\documentclass{jfm}

\usepackage{graphicx}
\usepackage{epstopdf, epsfig}
\usepackage[dvipsnames]{xcolor}
\usepackage{ulem}
\long\def\comment#1{}

\shorttitle{Rayleigh-Taylor instability during impact cratering experiments}
\shortauthor{V. Lherm, R. Deguen, T. Alboussi\`{e}re and M. Landeau}

\title{Rayleigh-Taylor instability during impact cratering experiments}

\author{
    V. Lherm\aff{1}\corresp{\email{victor.lherm@ens-lyon.fr}},
    R. Deguen\aff{2},
    T. Alboussi\`{e}re\aff{1}
    \and M. Landeau\aff{3}
  }

\affiliation{
\aff{1}Univ Lyon, ENSL, Univ Lyon 1, CNRS, LGL-TPE, F-69007 Lyon, France
\aff{2}Universit\'{e} Grenoble Alpes, CNRS, ISTerre, F-38041 Grenoble, France
\aff{3}Institut de Physique du Globe de Paris, CNRS, Universit\'{e} de Paris, Paris, France
}

\begin{document}

\maketitle

\begin{abstract}
When a liquid drop strikes a deep pool of a second liquid, an impact crater opens while the liquid of the drop decelerates and spreads on the surface of the crater. If the density of the drop is larger than the surrounding, the interface between the drop liquid layer and its surrounding becomes unstable, producing mushroom-shaped plumes growing radially outward.
We interpret this instability as a spherical Rayleigh-Taylor instability associated with the deceleration of the interface between the drop and its surrounding, which significantly exceeds the ambient vertical gravity. We investigate experimentally how changing the density contrast and the impact Froude number affects the instability and the resulting mixing layer.
Using backlighting and planar laser-induced fluorescence methods, the position of the air-liquid interface, the thickness of the mixing layer, and an estimate of the instability wavelength are obtained.
First, the evolution of the mean crater radius is derived from an energy conservation model. The observed mixing layer dynamics is then explained by a model involving a competition between the geometrical expansion of the crater tending to decrease its thickness, and mixing produced by the Rayleigh-Taylor instability tending to increase its thickness. The estimated instability wavelength is finally compared to an approximate linear stability analysis of a radially accelerated fluid sphere into a less dense fluid.
Mixing properties of this impact-related instability have geophysical implications regarding the differentiation of terrestrial planets, in particular by estimating the mass of magma ocean silicates that equilibrates with the metal core of the impacting planetesimals.
\end{abstract}

% \tableofcontents %TO REMOVE

\section{Introduction}
\label{sec:introduction}

%RTI DEFINITION
The Rayleigh-Taylor (RT) instability refers to the perturbation of a horizontal interface between two fluids of different densities.
In a gravitational field, an interface separating a dense fluid supported by a lighter one is unstable \citep{rayleigh_1899}. In this static case, the average position of the interface does not vary with time.
If the interface is accelerated in the direction from the lighter to the denser fluid, the configuration is also unstable \citep{taylor_1950}. In this dynamic case, the average position of the interface varies in time.
In both cases, infinitesimal perturbations at the interface will grow in time, leading to the interpenetration of the fluids, and to the reduction of their combined potential energy.

%SPHERICAL INTERFACES
The RT instability was first investigated at planar interfaces using theoretical, numerical, and experimental methods, both in the early-time linear \citep[\textit{e.g.}][]{emmons_1960,chandrasekhar_1961,tryggvason_1988} and the subsequent non-linear regimes \citep[\textit{e.g.}][]{linden_1994,dalziel_1999,dimonte_1999}.
However, various phenomena such as inertial confinement fusion (ICF) experiments \citep[\textit{e.g.}][]{lindl_1998,thomas_2012}, supernovae explosions \citep[\textit{e.g.}][]{arnett_1989,schmidt_2006}, detonation of explosive charges \citep[\textit{e.g.}][]{balakrishnan_2011}, and collapsing bubbles  \citep[\textit{e.g.}][]{prosperetti_1977,lin_2002}, involve RT instabilities at spherical interfaces.
The spherical configuration was initially investigated in static and dynamic cases, regarding the early-time linear stability of spherical interfaces between two inviscid fluids  \citep{bell_1951,plesset_1954,mikaelian_1990}. Viscosity effects responsible for energy dissipation at small-scale were also investigated in both cases \citep{chandrasekhar_1955,prosperetti_1977,mikaelian_2016}.
Turbulent mixing related to the late-time non-linear RT instability dynamics was also investigated for spherical interfaces \citep{youngs_2008,thomas_2012,lombardini_2014}.

\begin{figure}
    \centering
    \includegraphics[width=1\linewidth]{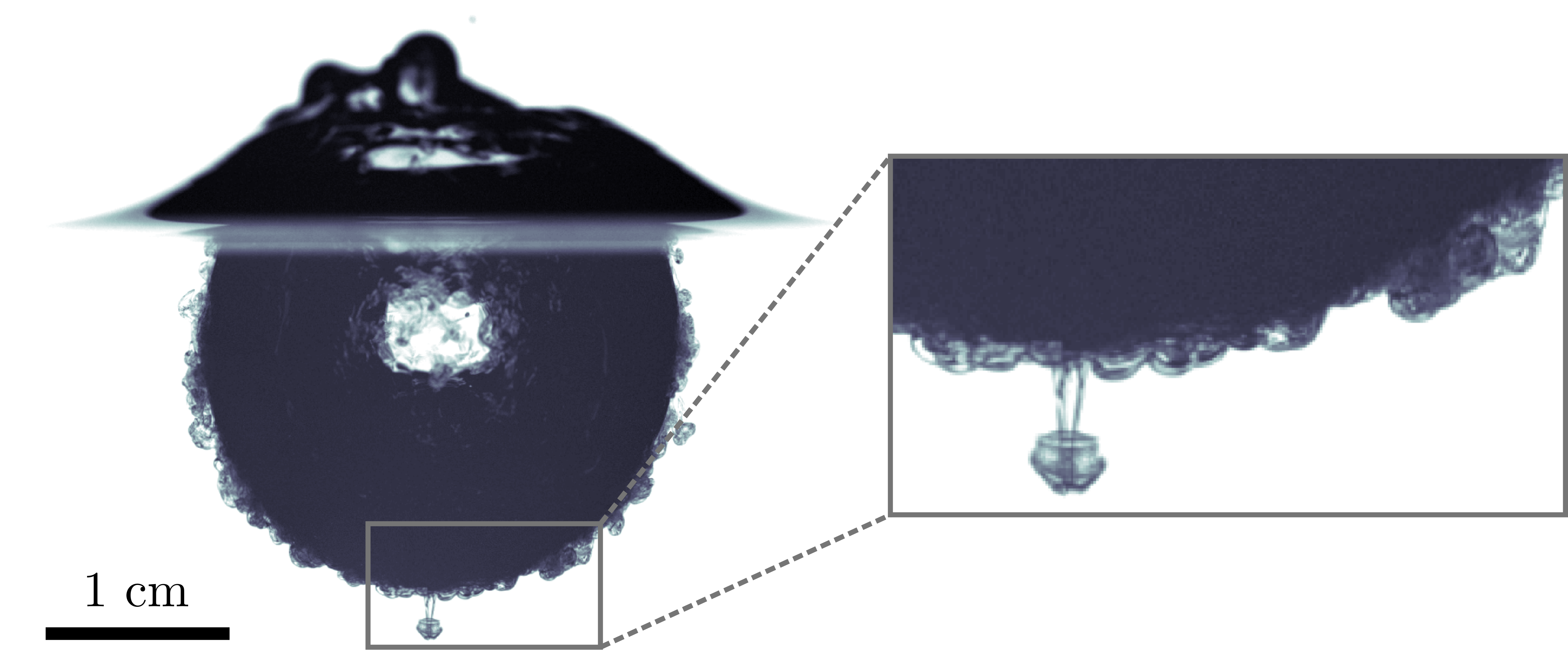}
    \caption{Crater produced by the vertical impact of a liquid drop onto a less dense liquid pool. A spherical Rayleigh-Taylor instability develops around the crater when it decelerates, which results in the mushroom-shaped plumes growing radially outward.}
    \label{fig:RT_intro}
\end{figure}

%DROP IMPACT
A RT instability at spherical interfaces is also expected to occur during drop impact, which is the focus of the experiments discussed in this paper. When a liquid drop strikes a less dense deep liquid pool, the crater opening deceleration produces an unstable equilibrium where perturbations at the drop-pool interface are amplified (figure \ref{fig:RT_intro}).
The RT instability dynamics depends crucially on the acceleration history of the interface \citep{mikaelian_1990,dimonte_2000}. In the case of a drop impact, acceleration history is dictated by the crater opening dynamics, which depends on the impact parameters: drop radius, impact velocity, ambient gravity and physical properties of the fluids such as surface tension, density and viscosity.
Depending on these impact parameters, various impact regimes such as bouncing, coalescence and splashing may occur \citep[\textit{e.g.}][]{rein_1993}. Given the impact energy of the drops involved, all our experiments are in the splashing regime. 
Since the pioneering experiments of \citet{worthington_1895}, the splashing regime has been extensively investigated \citep{engel_1966,engel_1967,macklin_1976,pumphrey_1990,prosperetti_1993,morton_2000,leng_2001,fedorchenko_2004,bisighini_2010,ray_2015}. In particular, the effects of immiscibility \citep{lhuissier_2013,jain_2019}, viscoplasticity \citep{jalaal_2019}, impact angles \citep{okawa_2006,gielen_2017}, and thickness of the target layer \citep{berberovic_2009} on impact dynamics have been examined.
Based on these experiments, several models of the crater size evolution and the related acceleration history were developed, using energy conservation \citep[\textit{e.g.}][]{engel_1966,engel_1967} or momentum conservation in an irrotational flow \citep[\textit{e.g.}][]{bisighini_2010}.

%GEOPHYSICAL APPLICATIONS
Besides providing an example of a RT instability at a spherical interface, this drop impact instability and the mixing related to it have geophysical implications.
Terrestrial planets such as the Earth formed 4.5 billions years ago by the successive accretion of increasingly massive bodies composed mainly of silicates and iron \citep{chambers_2010}, the last giant impact being probably responsible for the formation of the moon \citep{canup_2012,cuk_2012}. During this accretion process, planetary materials are heated by the kinetic energy released during the impacts, the reduction of gravitational potential energy as the metal of the impactors migrates toward the core, and the decay of radioactive isotopes \citep{rubie_2015}. This energy supply contributes to the production of deep magma oceans \citep{solomatov_2015}. In addition, accretion models show that most of the Earth mass was accreted from differentiated bodies, \textit{i.e.} with a separate core and mantle \citep{kleine_2002,schersten_2006}. Both the impacting body and the planetary surface are melted by the shock-waves produced by the impact, releasing the liquid metal core of the impactor into a fully-molten magma ocean \citep{tonks_1993}. This results in a situation where the metal core of an impactor strikes a less dense silicate magma ocean. A spherical RT instability is then expected to occur during crater opening, producing mixing that contributes to the thermal and chemical equilibration between the metal core of the impactors and the silicates of the magma ocean.

The current dynamics of the Earth is partly inherited from its concomitant accretion and differentiation. Heat partitioning and chemical fractionation between the mantle and the core depend on the physical processes involved during differentiation \citep{stevenson_1990,wood_2006}, which includes in particular equilibration and dispersion occurring during planetary impacts \citep{canup_2004,cuk_2012,kendall_2016,landeau_2020,nakajima_2020}.
Heat partitioning sets the initial temperature contrast between the mantle and the core. It crucially determines the early thermal and magnetic evolution of the planet, in particular the formation and evolution of magma oceans \citep{labrosse_2007,sun_2018}, and the existence of an early dynamo \citep{williams_2004,monteux_2011,badro_2018}. 
Chemical fractionation has also major geodynamical implications, such as the nature and abundance of radioactive and light elements in the core \citep{corgne_2007,siebert_2012,badro_2015,fischer_2015}.
Geochemical data such as isotopic ratios and partitioning coefficients between metal and silicates provide constraints on the timing of accretion and physical conditions of core formation in terrestrial planets \citep{li_1996,kleine_2002,righter_2011,siebert_2011}. However, their interpretation depends on the degree of chemical equilibration between the metal of the impactors' core and the magma ocean \citep{rudge_2010,rubie_2011}.

Consequently, an estimate of the mixing produced by the spherical RT instability during the impact is required in order to properly interpret geochemical data. In this paper, we examine the spherical RT instability produced during an impact using fluid dynamics experiments. Concerning planetary impacts, experiments allow to capture small-scale processes crucial to mixing quantification, whereas numerical simulations, more realistic on a large scale (sphericity, angled impact, self-gravitation), fail to capture these processes due to their resolution limit typically around 10 km \citep[\textit{e.g.}][]{kendall_2016}.

%OUTLINE
After a phenomenological description of the RT instability, the crater radius evolution is obtained using an energy conservation model. Knowing the acceleration history of the cratering process, the mixing layer dynamics is characterised, regarding in particular the evolution of the mixing layer thickness and the early-time instability wavelength. Mixing properties of this impact-related instability are eventually applied to the differentiation of terrestrial planets, in particular by estimating the mass of silicates that equilibrates with the metal of the impactors during crater opening, in particular during potential Moon-forming impacts.

\section{Impact cratering experiments}
\label{sec:experiments}

\subsection{Dimensional analysis}

We expect that the impact dynamics of a liquid drop released above a deep liquid pool with a different density and viscosity depends on its impact velocity $U_i$ and radius $R_i$, the densities $\rho_1$ and $\rho_2$ of the drop and the pool, the dynamic viscosities $\mu_1$ and $\mu_2$ of the drop and the pool, the surface tension at the air-liquid interface $\sigma$, and the acceleration of gravity $g$.
Since these eight parameters contain three fundamental units, the Buckingham-Pi theorem dictates that the impact dynamics depends on a set of five independent dimensionless numbers. We choose the following:
\begin{equation}
    Fr=\frac{U_i^2}{g R_i}, \quad
    We=\frac{\rho_1U_i^2R_i}{\sigma}, \quad
    Re=\frac{\rho_2 U_i R_i}{\mu_2}, \quad
    \rho_1/\rho_2, \quad
    \mu_1/\mu_2.
    \label{eq:dimensionless_numbers}
\end{equation}
The Froude number $Fr$ is a measure of the relative importance of impactor inertia and gravity forces. It can also be interpreted as the ratio of the kinetic energy $\rho_1 R_i^3 U_i^2$ of the impactor to its gravitational potential energy $\rho_1 g R_i^4$ just before impact. The Weber number $We$ compares the impactor inertia and interfacial tension at the air-liquid interface. The Reynolds number $Re$ is the ratio between inertial and viscous forces. $\rho_1/\rho_2$ and $\mu_1/\mu_2$ compare respectively the density and the dynamic viscosity of the drop and the pool. Since surface tension depends on salt concentration, a surface tension ratio between the drop and the pool is also involved. However, we neglect this parameter because the Weber number is much larger than unity and because the surface tension of the drop only varies by a maximum of 20\% compared to the pool. We will also make use of a modified Froude number and the Bond number,
\begin{equation}
    Fr^*=\frac{\rho_1}{\rho_2}\frac{U_i^2}{g R_i}, \quad
    Bo=\frac{\rho_2 g R_i^2}{\sigma},
    \label{eq:Froude_star}
\end{equation}
which can respectively be understood as the ratio of the kinetic energy of the impactor $\rho_1 R_i^3 U_i^2$ to the change of potential gravitational energy $\rho_2 g R_i^4$ associated with the opening of a crater of size $R_i$, and the ratio of buoyancy forces to interfacial tension at the air-liquid interface.
Table \ref{tab:parameters} compares the value of these dimensionless parameters in the experiments and in planetary impacts.

\begin{table}
    \centering
    \begin{tabular}{ccc}
        Dimensionless number & Experiments & Planetary impacts \\
        \hline
        $Fr$ & $60 - 1200$ & $1 - 10^5$\\
        $Fr^*$ & $60 - 2100$ & $1 - 10^5$\\
        $We$ & $60 - 1300$ & $\gtrsim 10^{14}$\\
        $Bo$ & $0.7 - 1$ & $\gtrsim 10^{10}$\\
        $Re$ & $2500 - 13500$ & $\gtrsim 10^{11}$\\
        $\rho_1/\rho_2$ & $1 - 1.8$ & 2\\
        $\mu_1/\mu_2$ & $0.9 - 1.2$ & 0.1
    \end{tabular}
    \caption{Typical values of the main dimensionless parameters (equations \ref{eq:dimensionless_numbers} and \ref{eq:Froude_star}) in the experiments, and typical planetary impacts. For planetary impacts, dimensionless numbers use a density of $4000~\mathrm{kg.m^{-3}}$ for molten silicates, and of $8000~\mathrm{kg.m^{-3}}$ for molten metal, a dynamic viscosity of $0.1~\mathrm{Pa.s}$ for molten silicates, and of $0.01~\mathrm{Pa.s}$ for molten metal \citep{solomatov_2015}. Surface tension between air and molten silicates, and between air and molten metal, are typically $0.3~\mathrm{J.m^{-2}}$ \citep{taniguchi_1988} and $1.8~\mathrm{J.m^{-2}}$ \citep{wille_2002}, respectively. Impact velocity is assumed to be one to three times the escape velocity \citep{agnor_1999,agnor_2004}. Impactor to target radius ratio is assumed to be in the range $10^{-4}-1$.}
    \label{tab:parameters}
\end{table}

Since experimental Reynolds numbers $Re \gtrsim 2500$ and Weber numbers $We \gtrsim 60$ are larger than unity, viscosity and surface tension are mostly negligible during crater opening.
Although $Re$ and $We$ are much larger during planetary impacts than in our experiments, this means that the cratering process and the RT instability are governed by inertia and buoyancy forces, in both our experiments and planetary impacts.
We thus focus on a regime, sometimes called gravity regime \citep{melosh_1989}, where the dynamics depends mainly on two dimensionless parameters, the Froude number $Fr$ and the density ratio $\rho_1/\rho_2$.

In order to characterise the cratering dynamics and the RT instability following the impact, we vary the drop radius, drop density and impact velocity.
We obtain Froude numbers and modified Froude numbers larger than unity, in the range $Fr \simeq 60-1200$ and $Fr^* \simeq 60-2100$, respectively. 
During planetary impacts, the Froude number (equation \ref{eq:Fr_planet}) is about 1 for impactors comparable in size with the target, but increases by several order of magnitude for small colliding bodies, \textit{e.g.} the Froude number is about $10^4$ for a 1 km radius body impacting an Earth-sized planet. Our experiments thus typically match planetary target to impactor radius ratio in the range $30-600$. 
We focus on five density ratios $\rho_1/\rho_2 \simeq \left\{1.0,1.2,1.4,1.6,1.8\right\}$, allowing a quantitative investigation of the density effects, in comparison with a reference case without density contrast.
We have also made a few experiments at $\rho_1/\rho_2 \simeq 0.8$ using ethanol in the drop.
During planetary impacts, the density ratio is expected to be about 2, which is close to the upper limit of our experimental density ratios.

During planetary impacts, several effects neglected in our experiments may change the dynamics of the cratering process and the related RT instability.
First, immiscibility between metal and silicates is not present in our experiments with miscible fluids. However, this effect may be neglected since inertial forces exceed widely interfacial forces in planetary impacts.
Then, the Mach number, defined as the ratio between the impact velocity and the sound velocity (P-wave velocity) in the magma ocean, is expected to be larger than unity during a planetary impact \citep[\textit{e.g.}][]{stixrude_2009}, whereas it is typically smaller than $3 \times 10^{-3}$ in our experiments. Compressibility effects such as the conversion of kinetic energy into heating are neglected.
Finally, the viscosity ratio between metal and silicates is also smaller than the experimental viscosity ratio by one order of magnitude. Although in this regime viscosity does not participate in the crater size evolution, it remains nonetheless a crucial parameter regarding the RT instability wavelength.

\subsection{Experimental set-up}

\subsubsection{Drop production, fluids, and cameras}

\begin{figure}
    \centering
    \includegraphics[width=1\linewidth]{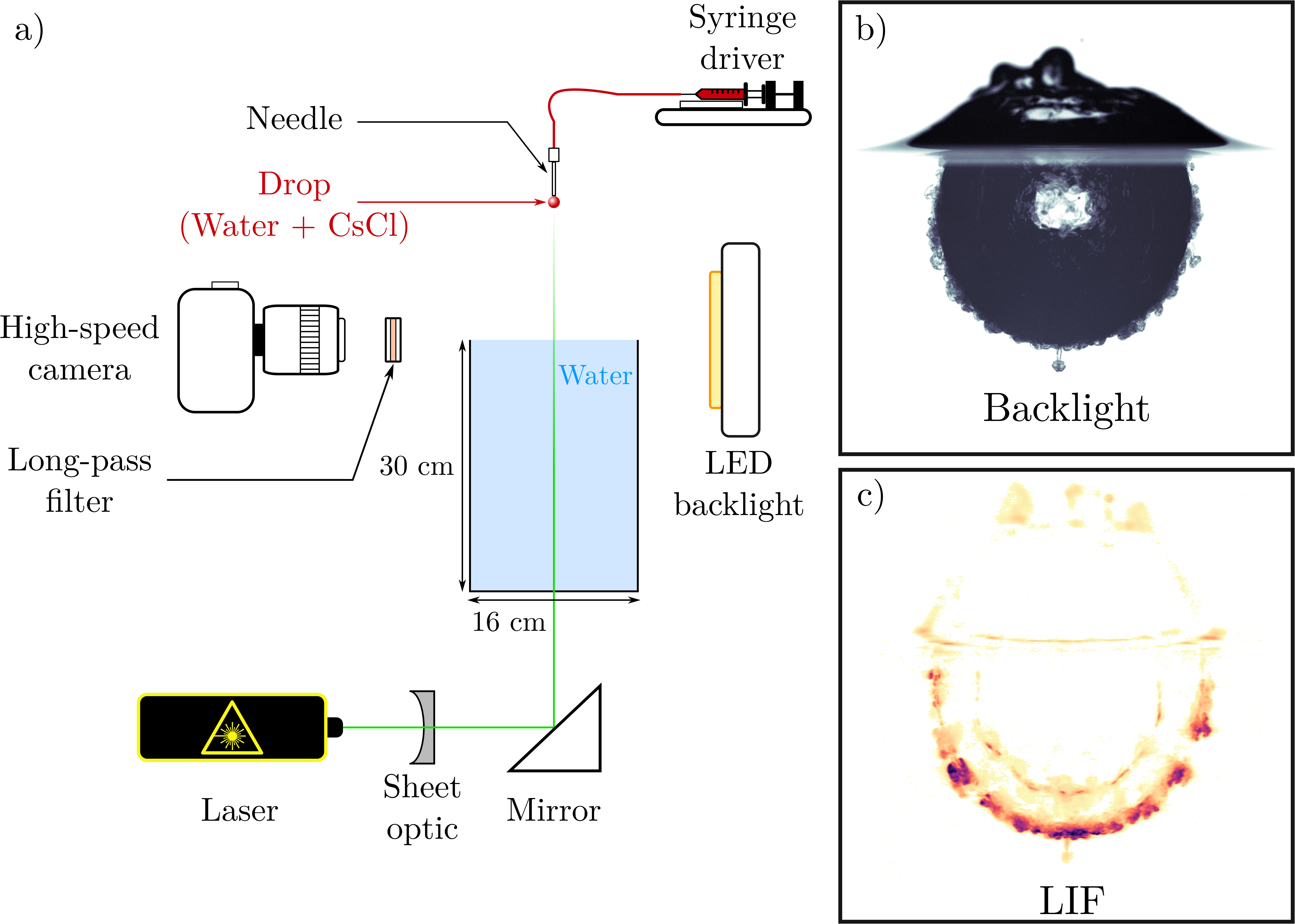}
    \caption{(a) Schematic view of the experimental set-up, including backlight and LIF configuration set-up. (b) Snapshot obtained using the backlight configuration. (c) Snapshot obtained using the LIF configuration.}
    \label{fig:setup}
\end{figure}

In these experiments, a liquid drop is released in the air above a deep liquid pool contained in a $16\times16\times30$ cm glass tank (figure \ref{fig:setup}). The pool level is exactly set at the top of the tank. The aim is to minimise the thickness of the meniscus on the side of the tank in order to obtain an image of the crater all the way to the surface. 

The drop is generated using a needle supplied with fluid by a syringe driver at a slow and steady pace. When the weight of the drop exceeds the stabilising surface tension effect, the drop comes off. We used a metallic needle with an inner diameter of 1.6 mm and a nylon plastic needle with an inner diameter of 4.7 mm, generating two series of drop size with typical radius in the range $1.7-2.0$ mm and $2.3-2.7$ mm, respectively. The exact drop size, which depends on the drop density, is calculated for each experiment based on a careful calibration using precise mass measurements of dozens of drops, independent density measurement, and assuming the drop is spherical. We validate this method using high-speed pictures of the drop prior to impact where we can directly measure the drop radius.

Typical impact velocities are in the range $1-5~\mathrm{m.s^{-1}}$. Impact velocity is calculated for each experiment using a calibrated free fall model for the drop including a quadratic drag. We also validate this method using high-speed pictures of the drop prior to impact where we can directly measure the drop velocity.

Concerning fluids, we use an aqueous solution of caesium chloride CsCl ($\rho_1=998-1800~\mathrm{kg.m^{-3}}$, $\mu_1=0.9 \times 10^{-3}-1.2 \times 10^{-3}~\mathrm{Pa.s}$) in the drop, and water ($\rho_2=998~\mathrm{kg.m^{-3}}$, $\mu_2=10^{-3}~\mathrm{Pa.s}$) in the pool. Surface tension at the air-water interface is $\sigma=73~\mathrm{mJ.m^{-2}}$. The density is measured for each experiments using an Anton Paar DMA 35 Basic densitometer. Viscosities and surface tension are obtained using data from \citet{haynes_2016}.

Results are obtained with two imaging configurations, backlight and Laser-Induced Fluorescence (LIF) configurations, most suited to crater shape determination and mixing characterisation, respectively. In both configurations, the camera is positioned at the same height as the water surface. Images are recorded at 1400 Hz with a $2560\times1600$ pixels resolution, and a 12 bits dynamic range, using a high-speed Phantom VEO 640L camera and a Tokina AT-X M100 PRO D Macro lens.

\begin{figure}
    \centering
    \includegraphics[width=1\linewidth]{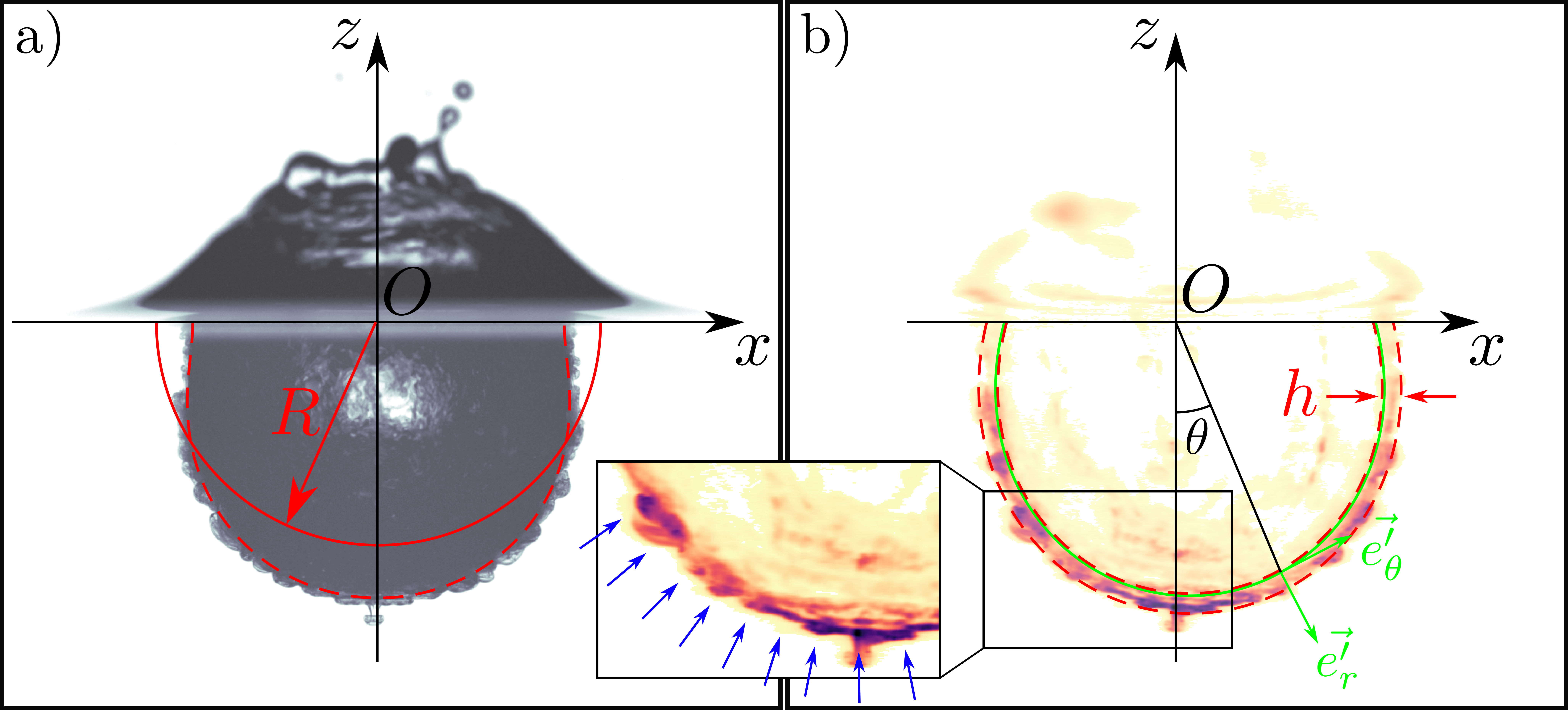}
    \caption{(a) Detection of the crater boundary. The dashed line corresponds to the fitted crater boundary using a set of shifted Legendre polynomials up to degree $l=5$, the degree $l=0$ gives the solid line corresponding to the mean crater radius. (b) Detection of the mixing layer. The solid line corresponds to the fitted crater boundary using a set of Legendre polynomials up to degree $l=5$. Dashed lines correspond to the weighted average mixing layer boundaries, calculated using the second moment of the concentration about the mean position of the layer. Blue arrows indicate the position of the plumes produced by the instability.}
    \label{fig:measure}
\end{figure}

\subsubsection{Backlight configuration}

In the backlight configuration (figure \ref{fig:measure}a), we measure the evolution of the mean crater radius. The crater is illuminated from behind by a LED backlight panel and appears dark owing to refraction of light at the air-water interface.
Image processing involves spatial calibration using a sight, background removal, intensity threshold, image binarization, and allows us to determine the crater boundary.

The crater boundary radius $R(\theta,t)$, which depends on the polar angle $\theta$ and time $t$, is then fitted using a set of shifted Legendre polynomials $\tilde{P}_l$ up to degree $l=5$
\begin{equation}
    R(\theta,t)=\sum_{l=0}^5 a_l\tilde{P}_l(\cos\theta),
\end{equation}
where $a_l$ are the fitted coefficient. The shifted Legendre polynomials are defined as an affine transformation of the standard Legendre polynomials $\tilde{P}_l(x)=P_l(2x-1)$, and are orthogonal on $[0,1]$, \textit{i.e.} on a half-space. The mean crater radius is obtained from the $l=0$ coefficient, \textit{i.e.} $R=a_0$. 

Uncertainties are dominated by the extrinsic variability between experiments in the same configuration. Each experiment is repeated at least four times consecutively in order to estimate uncertainties on dimensionless parameters and target quantities. This allows to include uncertainties resulting from reflections and refraction at the crater boundary.

In comparison, intrinsic uncertainties related to the spatial resolution of the camera, the spatial calibration, and image processing are negligible. Spatial resolution of the camera, \textit{viz} $30~\mathrm{px.mm^{-1}}$, is adequate given the size of the target, allowing this uncertainty to be neglected. Spatial calibration errors, typically around 0.2 px, are also neglected. Given the camera resolution and dynamic range, a good contrast is obtained on the crater and the impacting drop, which allows to neglect errors related to image processing.

Uncertainties on fluid properties and impact parameters are propagated to uncertainties on the dimensionless numbers (equations \ref{eq:dimensionless_numbers} and \ref{eq:Froude_star}).
Errors on density, viscosity and surface tension are carefully measured (density) or calculated (viscosity and surface tension) in a temperature controlled environment. Errors on the velocity and radius of the impacting drop are obtained from the variability in mass measurements and from error propagation in the velocity model, respectively.

\subsubsection{Laser-Induced Fluorescence configuration}

In the LIF configuration (figure \ref{fig:measure}b), we measure the thickness of the mixing layer and the number of plumes produced by the RT instability. A vertical laser sheet (532 nm) excites the fluorescent dye (Rhodamine 6G) contained in the fluid of the drop. The fluorescent dye then re-emits light between 570 nm and 660 nm. This emission signal is then recorded by the camera and isolated from the laser signal with a long-pass filter ($>540$ nm).
The laser sheet is generated using a 10 W Nd:YAG continuous laser in combination with a divergent cylindrical lens and a telescope, producing a 1 mm thick sheet. The laser sheet is diverted vertically using a $45^\circ$ mirror beneath the tank.
In order to isolate the mixing layer, images are processed with spatial calibration using a sight, background removal, and laser sheet corrections, removing sheet inhomogeneities. Artefacts due to reflections on the surface and on the air-water interface are then filtered and removed. In particular, the internal reflection of the mixing layer (\textit{e.g.} figure \ref{fig:setup}c) is carefully removed.
The dye concentration field in the mixing layer is eventually obtained, using its direct proportionality to the measured field of light intensity.

As for the backlight configuration, the position of the crater boundary $R(\theta,t)$, corresponding to the inner boundary of the mixing layer, is fitted using a set of shifted Legendre polynomials up to degree $l=5$. A local frame of reference $(\boldsymbol{e_r'},\boldsymbol{e_\theta}')$ is then defined, where $\boldsymbol{e_r'}$ is normal to the fitted crater boundary and $\boldsymbol{e_\theta'}$ is tangent to it.
For each polar position $\theta$ about the crater boundary, the local mean position of the mixing layer $\langle r' \rangle$ is calculated, using the position of the pixels in the local frame of reference $(r',\theta)$ and the corresponding concentration field $c$
\begin{equation}
    \langle r' \rangle(\theta)=\frac{\int r' c(r',\theta)\mathrm{d}r'}{\int c(r',\theta)\mathrm{d}r'}.
\end{equation}
The local standard deviation $\sigma_{r'}$ about the local mean position of the mixing layer is then calculated
\begin{equation}
    \sigma_{r'}(\theta)=\sqrt{\frac{\int \left[r'-\langle r' \rangle(\theta)\right]^2 c(r',\theta)\mathrm{d}r'}{\int c(r',\theta)\mathrm{d}r'}}.
\end{equation}
The mixing layer thickness $h$ is eventually obtained with
\begin{equation}
    h=\frac{\int_{-\pi/2}^{\pi/2} 2\sigma_{r'}R^2|\sin\theta |w(\theta)\mathrm{d}\theta}{\int_{-\pi/2}^{\pi/2} R^2|\sin\theta |w(\theta)\mathrm{d}\theta},
\end{equation}
using a weighted average where $w=1/[1+\exp\{k(|\theta|-\theta_0)\}]$ is a symmetric logistic weight function whose steepness is $k=30$ and sigmoid's midpoint is $\theta_0=\pi/3$.
The logistic function allows to give more weight to the bottom of the crater, between $\theta=0$ and $\theta=\theta_0$, and less to the top of the crater, close to $\theta=\pm \pi/2$.
The use of such a weight function is motivated by the polar dependency of the LIF signal quality. Close to the surface, \textit{i.e.} at $\theta=\pm \pi/2$, signal intensity is reduced and imaging of the mixing layer undergoes significant perturbations, leading to a poor estimate of its extent. The crater is indeed illuminated from below, so that the laser sheet undergoes absorption as it goes through a dyed layer, and refraction as it goes through a layer with a variable density and index of refraction.
Imaging of the mixing layer close to the surface may also be perturbed directly by the air-water interface, causing reflection of the laser sheet.
All these effects can be amplified since the crater is not hemispherical. If the drop is denser than the pool, the crater is stretched downward, leading after a while to an ellipsoidal crater centred below the surface of the pool. The path of the laser sheet through the mixing layer is thus geometrically increased, and is more likely to cross the air-water interface.

The number of plumes produced by the RT instability is counted manually for each experiment at the same dimensionless time $t/(R_i/U_i) \sim 10$  (figure \ref{fig:measure}b). It corresponds to an already developed and visible instability, where the plumes, however, did not have time to interact with each other, which is relevant since this number of plume is to be compared with theoretical results from a linear stability analysis \citep{chandrasekhar_1955}.

As in the backlight configuration, uncertainties are dominated by the extrinsic variability between experiments in the same configuration, and each experiment is thus repeated at least four times consecutively. Intrinsic uncertainties are negligible and uncertainties on dimensionless numbers are calculated in the same way.

\section{Experimental phenomenology}
\label{sec:phenomenology}

The phenomenology of the Rayleigh-Taylor instability is intrinsically related to the crater evolution following the impact, and particularly to its acceleration history. Using both backlight and LIF configurations, the air-water interface evolution and the mixing layer evolution are obtained, thus providing the means for a phenomenological description of the observed RT instability. The following description is based on two typical experiments, with and without density contrast, with the same crater opening dynamics, in the backlight (figure \ref{fig:sequence_01}a,c) and the LIF configurations (figure \ref{fig:sequence_01}b,d).

\subsection{Crater geometry}

\begin{figure}
    \centering
    \includegraphics[width=1\linewidth]{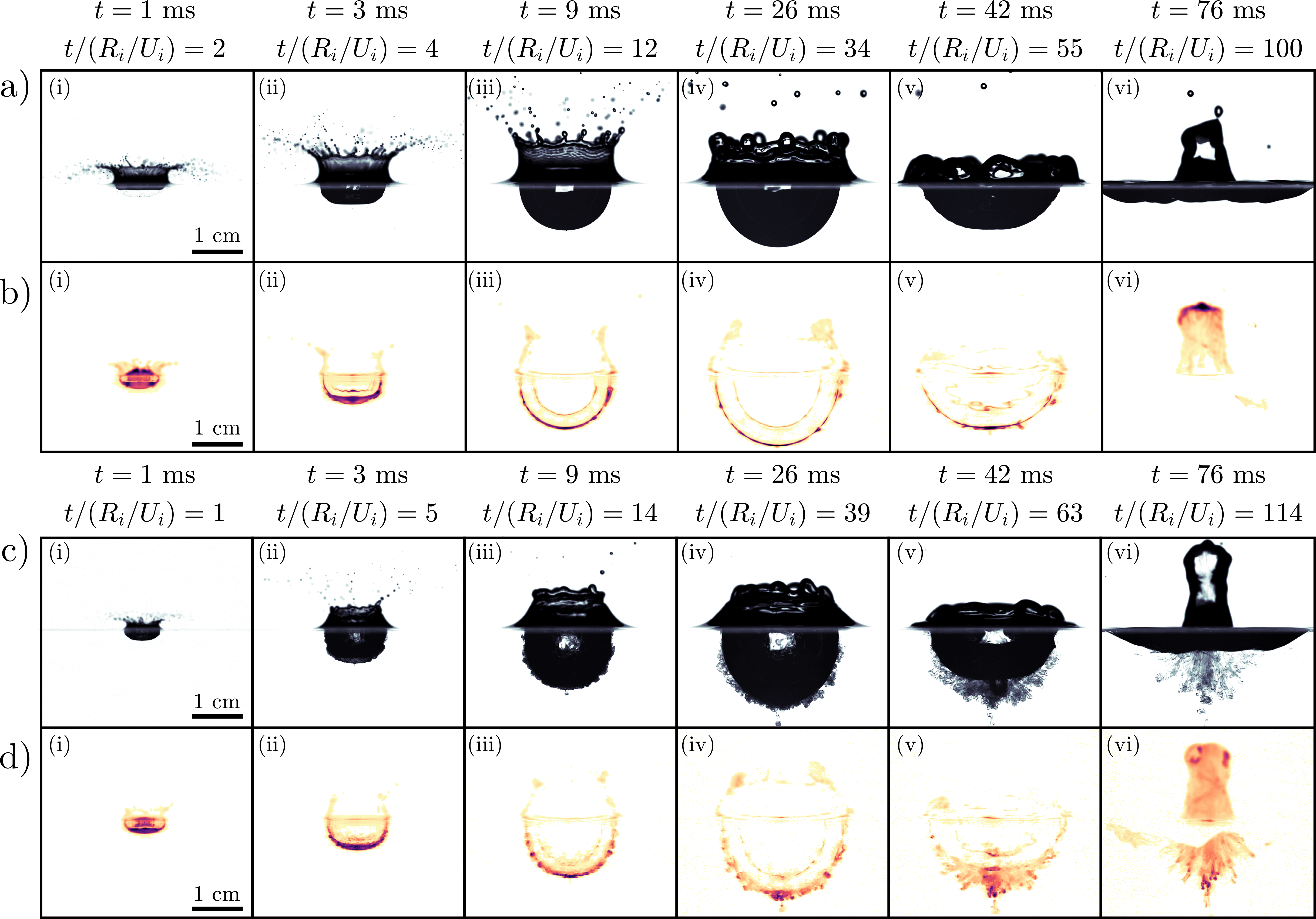}
    \caption{Liquid drop impact onto a deep liquid pool without density contrast ($\rho_1/\rho_2=1$ and $\mathrm{Fr}=481$) in the backlight (a) and the LIF (b) configuration. Liquid drop impact onto a deep liquid pool with density contrast  ($\rho_1/\rho_2=1.8$ and $\mathrm{Fr}=542$) in the backlight (c) and the LIF (d) configuration.}
    \label{fig:sequence_01}
\end{figure}

The impact of the drop causes the formation of an impact crater that grows until it reaches its maximum size (figure \ref{fig:sequence_01}a, iv). The liquid of the drop is first deformed and accumulated on the crater floor (figure \ref{fig:sequence_01}b, i), on a timescale $t/(R_i/U_i) \sim 2-3$, akin to previous results \citep{bisighini_2010}. Then, it quickly spreads on the crater sides toward the surface (figure \ref{fig:sequence_01}b, ii) during a timescale $t/(R_i/U_i) \sim 8$, eventually producing a layer with an approximately uniform thickness on the nearly hemispherical surface of the crater (figure \ref{fig:sequence_01}b, iii-v).
The impact also produces a fluid crown \citep{fedorchenko_2004} (figure \ref{fig:sequence_01}a, i-iv), along with a surface wave propagating radially outward from the crater \citep{leng_2001} on the horizontal surface. As can be seen from the concentration field (figure \ref{fig:sequence_01}b, i-iv), the fluid of the drop mostly accumulates on the surface of the crater, leaving a crown mainly composed of fluid from the pool. As soon as the crown decelerates, the cylindrical sheet produces liquid ligaments around the crown rim, which eventually fragment into drops (figure \ref{fig:sequence_01}a, ii-iii) \citep{krechetnikov_2009,zhang_2010,agbaglah_2013}.
When the crater reaches its maximum size (figure \ref{fig:sequence_01}a, iv), crater starts to collapse (figure \ref{fig:sequence_01}a, v). A capillary wave can develop on the crater surface \citep{pumphrey_1990,morton_2000}. The resulting converging flow leads to the formation of an upward jet mostly composed by the fluid of the drop, in view of the concentration field (figure \ref{fig:sequence_01}b, vi).

\subsection{Rayleigh-Taylor instability}

During crater opening, after an initial impulse, the crater boundary decelerates, producing a radial deceleration more than ten times larger than vertical gravity (see section \ref{sec:crater_size_time}). If the density of the drop is larger than the density of the pool, it leads to an unstable equilibrium where any perturbation at the drop-pool interface is amplified radially, in the local direction of apparent gravity. This density-driven instability is interpreted as a Rayleigh-Taylor instability since the (negative) acceleration history of the interface is sustained during the whole crater opening. Although this crater deceleration decreases rapidly in magnitude, the interface acceleration is not impulsive. Thus, the instability is not an incompressible Richtmyer-Meshkov instability \citep{richtmyer_1960,meshkov_1969,jacobs_1996}.
This spherical RT instability is expected to compete with the crater geometrical expansion, that dampens its development by stretching the dense layer. 

First, the instability goes through a stage where the perturbations of the interface are small in comparison with the radius of the crater and the wavelength of the instability (figure \ref{fig:sequence_01}d, i-ii). This initial stage is expected to occur very early in the crater opening sequence, as suggested by our experiments where the perturbations reach the same size as the instability wavelength, at $t/(R_i/U_i) \sim 10$.
This stage is not directly observed in our experiments because geometrical effects, that stretch and thin the dense layer, prevail over the RT instability. Fast vigorous crater expansion is indeed expected to dampens the RT instability.
Since the two fluids involved are miscible, \textit{i.e.} surface tension is zero, all wave numbers are expected to be unstable with respect to the RT instability \citep{chandrasekhar_1955}. However, owing to larger velocity gradients at large wave numbers, viscosity is responsible for the energy dissipation of short wavelengths. The growth rate of the instability then decreases as the inverse of the wave number \citep{chandrasekhar_1961}. Consequently, a mode of maximum instability depending on the acceleration history and impact parameters is expected to develop. This mode of maximum instability likely determines the typical number of plumes and the corresponding wavelength. 

At some point, geometrical effects produced by crater expansion loose intensity and become comparable with the RT instability. This coincides with a stage where the instability is strongly influenced by three-dimensional effects, leading to the formation of plumes below the hemispherical surface of the crater (figure \ref{fig:sequence_01}d, iii). As the RT instability grows toward a more turbulent layer, the mode of maximum instability is likely to be modified by non-linear interactions.
Plumes then start interacting with each other, producing a mixing layer (figure \ref{fig:sequence_01}d, iv). Interactions are expected to come from multi-mode perturbations interacting non-linearly with each other, and from small-scale Kelvin-Helmholtz instability produced at the side of the plumes \citep{cook_2004}.

\begin{figure}
    \centering
    \includegraphics[width=\linewidth]{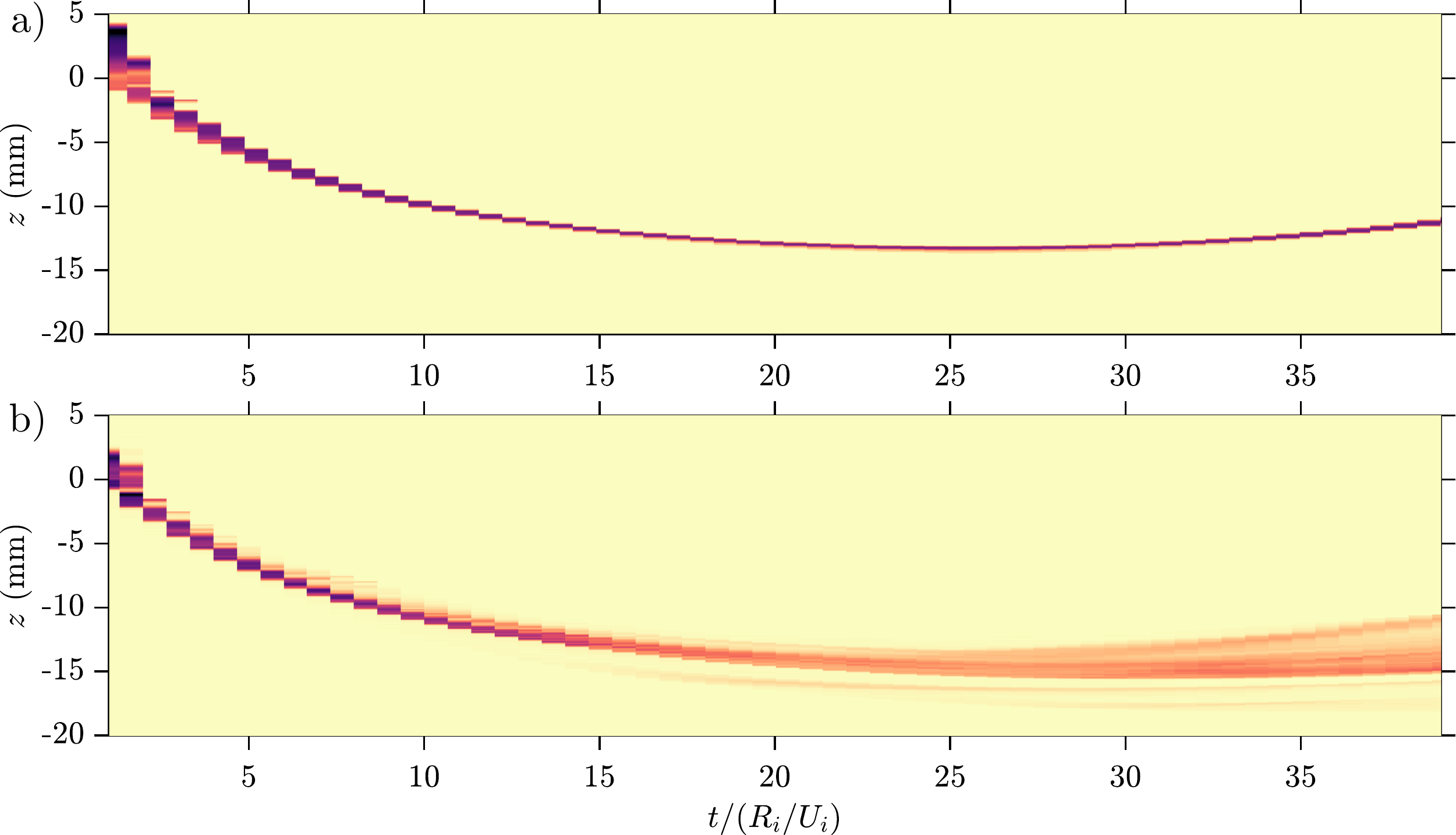}
    \caption{Position of the mixing layer cross-section at $\theta=0$, as a function of time, normalised by the drop free-fall time $R_i/U_i$, without (a) ($\rho_1/\rho_2=1.0$) and with (b) ($\rho_1/\rho_2=1.8$) density contrast between the drop and the pool. The surface of the pool prior to impact is at $z=0$. The colour scale represents the intensity of the fluorescent tracer initially within the drop.}
    \label{fig:h_time_cmp}
\end{figure}

A qualitative insight on the competition between the geometrical expansion of the crater and mixing produced by the RT instability is given with figure \ref{fig:h_time_cmp}, showing the time evolution of the position of the mixing layer during crater opening.
When the drop density is the same as in the pool (figure \ref{fig:h_time_cmp}a), no mixing by the RT instability occurs and the drop layer only becomes thinner due to crater expansion.
When the drop density is larger than the pool density (figure \ref{fig:h_time_cmp}b), the mixing layer also becomes thinner due to crater expansion, but at some point, the mixing layer thickness begins to increase as the mixing produced by the RT instability starts to prevail over the geometrical expansion of the crater.

\begin{figure}
    \centering
    \includegraphics[width=1\linewidth]{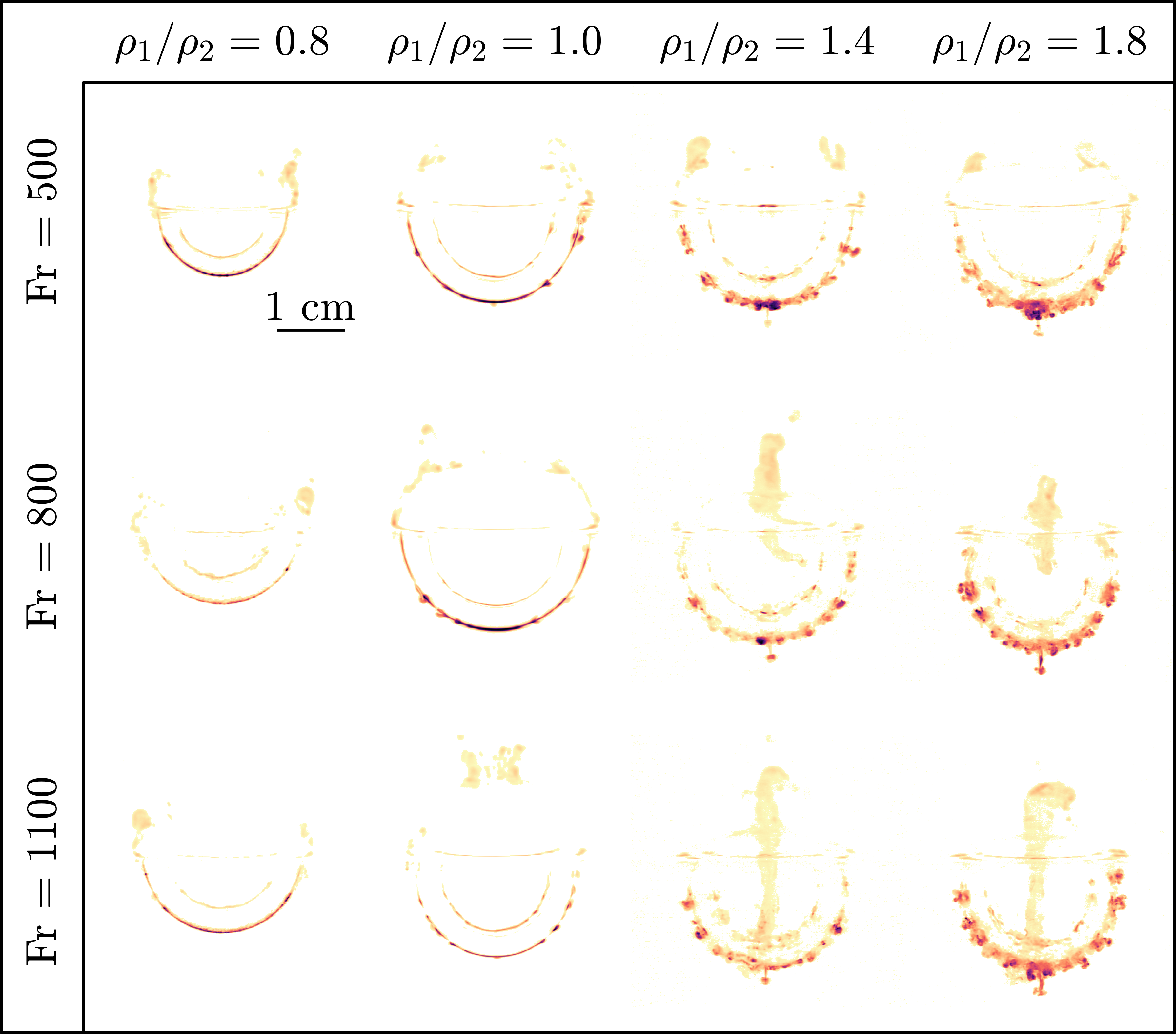}
    \caption{Snapshots of the mixing layer when the crater reaches its maximum size, as a function of the Froude number and the density ratio.}
    \label{fig:RT_regime_01}
\end{figure}

This competition also appears in figure \ref{fig:RT_regime_01}, which shows the mixing layers when the crater reaches its maximum size, as a function of the Froude number and the density ratio.
For density ratios smaller than unity (first column), the impacting drop is only stretched by crater expansion. Mixing related to the RT instability does not occur in this stable configuration.
For density ratios about unity (second column), mixing slightly occur due to large-scale Kelvin-Helmholtz instability in the mixing layer. During crater opening, the air-water interface is not purely hemispherical and the velocity field is not purely radial \citep{bisighini_2010}. This creates a velocity shear across the interface, and sometimes produces a large-scale Kelvin-Helmholtz instability.
For density ratios larger than unity (third and forth columns), mixing due to the RT instability occurs. For a given Froude number, the mixing layer thickness obtained when the crater reaches its maximum size increases with the initial density ratio. For a given density ratio, the mixing layer thickness does not change significantly with the Froude number.

\section{Evolution of the crater size}
\label{sec:crater_size}

Experiments in the backlight configuration provide the time evolution and the maximum of the mean crater radius, a required step in the understanding of the RT instability dynamics. We derive an energy model for the evolution of the crater radius, velocity, and acceleration and compare it with experiments. We then obtain scaling laws for the maximum crater radius and the crater opening timescale.

\subsection{Energy conservation model}

We use an energy conservation model \citep{engel_1966,engel_1967,leng_2001} accounting for the density difference between the impacting drop and the targeted pool. We assume a hemispherical crater associated with an incompressible and irrotational flow. Since the crater opening dynamics is mainly driven by impactor inertia and gravity forces ($Re \gtrsim 2500$), viscous dissipation is not included into the model. The terms related to the formation of the crown and the surface wave during the impact, in particular their potential, kinetic, and surface energies, are not explicitly included either in the model.

On the basis of these assumptions, the sum of the crater potential energy $E_p$, the crater surface energy $E_\sigma$, and the crater kinetic energy $E_k$, at any instant of time is equal to the impacting drop kinetic energy $E_i$ just before the impact.
The potential energy of the crater is
\begin{equation}
    E_p=\int\rho_2gz\mathrm{d}V=\int_0^R\rho_2g\pi\left(R^2-z^2\right)z\mathrm{d}z=\frac{1}{4} \pi \rho_2 g R^4,
\end{equation}
where $z$ is the depth.
The crater surface energy corresponds to the formation of a new surface due to crater opening. This energy is related to the difference between the initially planar surface area of the pool $\pi R^2$ and the hemispherical surface area of the cavity $2\pi R^2$, \textit{i.e.}
\begin{equation}
    E_\sigma=\sigma\left(2\pi R^2-\pi R^2\right)=\sigma\pi R^2.
\end{equation}
The crater kinetic energy corresponds to the kinetic energy of the pool fluid below the initial surface and is related to the flow velocity potential. A radial velocity potential of the form $\Phi=-A/r$, solution of the Laplace equation $\nabla^2\Phi=0$, is able to satisfy the boundary conditions. At the crater boundary, the radial velocity is $u_r(r=R)=(\partial\Phi/\partial r)_{r=R}=\dot{R}$, giving $A=\dot{R}R^2$ and
\begin{equation}
    \Phi=-\frac{\dot{R}R^2}{r}.
    \label{eq:potential}
\end{equation}
The radial velocity, the tangential velocity, and the resultant velocity are respectively
\begin{equation}
    \left\{
    \begin{array}{l}
        u_r=\frac{\dot{R}R^2}{r^2},\\
        u_\theta=0,\\
        ||\boldsymbol{u}||=\sqrt{u_r^2+u_\theta^2}=\frac{\dot{R}R^2}{r^2}.
    \end{array}
    \right.
    \label{eq:velocity}
\end{equation}
The crater kinetic energy is then
\begin{equation}
    E_k=\int\frac{1}{2}\rho_2||\boldsymbol{u}||^2\mathrm{d}V=\int_R^{+\infty}\pi\rho_2\dot{R}^2R^4\frac{1}{r^2}\mathrm{d}r=\pi\rho_2R^3\dot{R}^2.
\end{equation}
The impacting drop kinetic energy is
\begin{equation}
E_i=\frac{2}{3} \pi \rho_1 R_i^3 U_i^2.
\end{equation}

Energy conservation between $E_p$, $E_\sigma$, $E_k$ and $E_i$ gives
\begin{equation}
    \frac{1}{4} \rho_2 g R^4 + \sigma R^2 + \xi \rho_2 R^3 \dot{R}^2 = \frac{2}{3} \phi \rho_1 R_i^3 U_i^2,
\end{equation}
where $\phi$ and $\xi$ are fitted parameters.
The coefficient $\phi$ corresponds to a correction parameter accounting for the terms not included in the model, \textit{i.e.} viscous dissipation and crown energy terms. The coefficient $\xi$ is a correction parameter accounting for the difference between the deliberately simplified velocity potential used in the model and the true flow.

Normalising the crater radius and opening velocity by the impacting drop radius $R_i$ and velocity $U_i$, respectively, energy conservation becomes
\begin{equation}
    \frac{1}{4} \frac{1}{Fr^*}R^4 + \frac{1}{Fr^*Bo} R^2 + \xi\left(\frac{\rho_1}{\rho_2}\right)^{-1} R^3 \dot{R}^2 = \frac{2}{3}\phi.
    \label{eq:energy_model_adim}
\end{equation}

For each experiment, $\phi$ is calculated at $R=R_{max}$. Assuming that the velocity field vanishes simultaneously in the pool \citep{prosperetti_1993}, the crater kinetic energy vanishes when the crater reaches its maximum size, which gives an estimate of $\phi$ independent of the crater opening velocity field
\begin{equation}
    \phi=\frac{3}{2}\frac{1}{Fr^*}R_{max}^2\left(\frac{1}{4}R_{max}^2+\frac{1}{Bo}\right).
\end{equation}
Knowing the amount of energy delivered to the pool after the impact, the time evolution of the mean crater radius is then fitted to the experiments with equation \ref{eq:energy_model_adim} using a least-square method, the kinetic energy correction parameter $\xi$ being a fit parameter.
Knowing $\phi$ and $\xi$, the ordinary differential equation \ref{eq:energy_model_adim} is solved using the boundary condition $R(1)=1$. This condition assumes that the crater radius is initially the same as the drop radius, at $t=R_i/U_i$.

\subsection{Time evolution}
\label{sec:crater_size_time}

\begin{figure}
    \centering
    \includegraphics[width=1\linewidth]{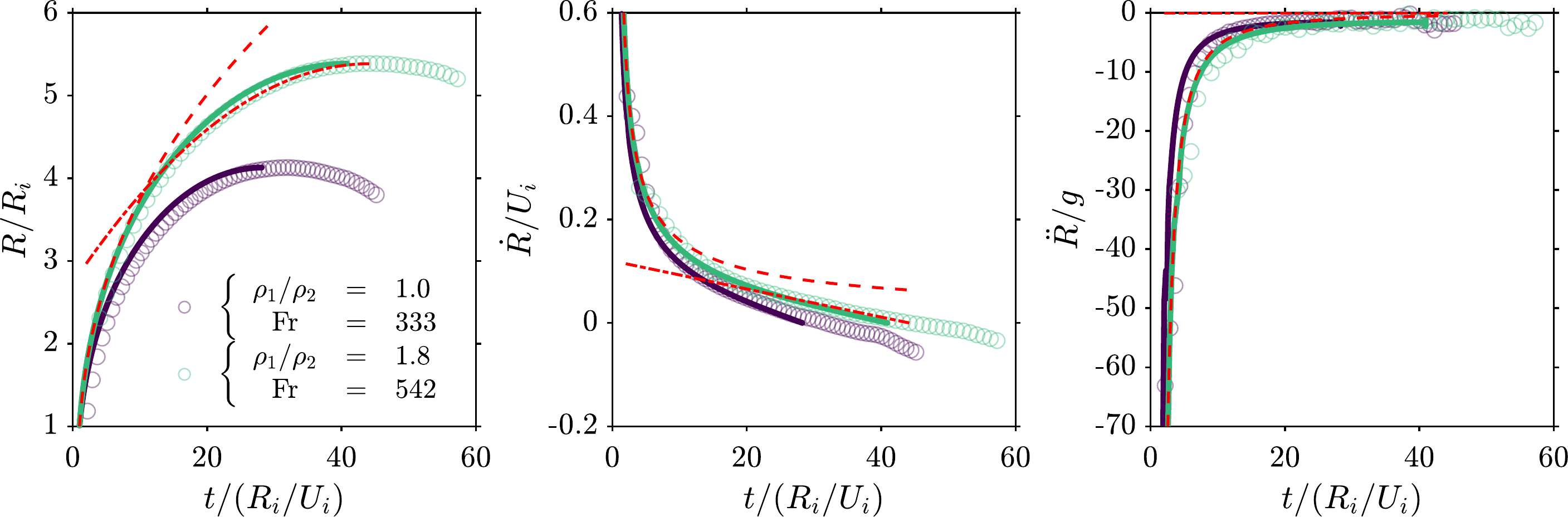}
    \caption{Time evolution, normalised by the drop free-fall time $R_i/U_i$, of the mean crater radius $R$ normalised by the drop radius $R_i$ (a), the mean crater velocity $\dot{R}$ normalised by the impact velocity $U_i$ (b), and the mean crater acceleration $\ddot{R}$ normalised by the acceleration of gravity $g$ (c), for two impact parameters. Circles and solid lines correspond respectively to experimental data and fitted energy model (equation \ref{eq:energy_model_adim}). Dashed lines and dashed-dotted lines correspond respectively to early-time power-law solution (equation \ref{eq:energy_model_assum}) and late-time quadratic solution (equations \ref{eq:quadratic_solution}) for the $\rho_1/\rho_2=1.8$ and $Fr=548$ experiment.}
    \label{fig:crater_size_01}
\end{figure}

Figure \ref{fig:crater_size_01} compares the fitted energy model with experimental data, in two reference cases, with and without density difference between the impacting drop and the pool. In both cases, the fitted mean crater radius, opening velocity, and acceleration are in close agreement with the experimental data. In the $\rho_1/\rho_2=1$ case, $\phi=0.40$ and $\xi=0.35$. In the $\rho_1/\rho_2=1.8$ case, $\phi=0.39$ and $\xi=0.34$.

At early times, the crater potential and surface energies may be neglected in comparison with the crater kinetic energy. The kinetic energy of the impactor is then balanced by the crater kinetic energy in equation \ref{eq:energy_model_adim}. Using these assumptions, a power-law scaling for the crater evolution is obtained
\begin{equation}
    \left\{
    \begin{array}{l}
        R(t)=\left[Q(t-1)+1\right]^{2/5}\\
        \dot{R}(t)=\frac{2}{5}Q\left[Q(t-1)+1\right]^{-3/5}\\
        \ddot{R}(t)=-\frac{6}{25}Q^2\left[Q(t-1)+1\right]^{-8/5}
    \end{array}
    \right.,
    \label{eq:energy_model_assum}
\end{equation}
where $Q=\left(\frac{25}{6}\frac{\phi}{\xi}\frac{\rho_1}{\rho_2}\right)^{1/2}$. This scaling consistently verifies the imposed boundary condition $R(1)=1$.
The scaling depends on the density ratio $\rho_1/\rho_2$, and on the correction parameters $\phi$ and $\xi$. It is in agreement with experimental data at early times (figure \ref{fig:crater_size_01}, dashed lines), and similar scalings from previous works \citep{leng_2001,bisighini_2010}.

At late times, the crater velocity becomes very small. If surface tension can be neglected, taking the time derivative of equation \ref{eq:energy_model_adim}, and then making the assumption $\dot{R}=0$ gives
\begin{equation}
    \ddot{R}=-\frac{1}{2}\frac{1}{Fr\xi}.
    \label{eq:quadratic_equation}
\end{equation}
Using $R(t_{max})=R_{max}$ and $\dot{R}(t_{max})=0$ as boundary conditions, a quadratic solution is obtained
\begin{equation}
        \left\{
    \begin{array}{l}
        \ddot{R}(t)=-\frac{1}{2}\frac{1}{Fr\xi}\\
        \dot{R}(t)=-\frac{1}{2}\frac{1}{Fr\xi}(t-t_{max})\\
        R(t)=R_{max}-\frac{1}{4}\frac{1}{Fr\xi}(t-t_{max})^2
    \end{array}
    \right..
    \label{eq:quadratic_solution}
\end{equation}
This scaling is in good agreement with experimental data at late times using experimental values for the boundary conditions (figure \ref{fig:crater_size_01}, dash-dotted lines).
Using scaling laws for $R_{max}$ and $t_{max}$ (determined in section \ref{sec:crater_size_maximum}), the late-time quadratic evolution of the mean crater radius may be fully predicted as function of $Fr$, $\phi$, $\xi$, and $\rho_1/\rho_2$.

\subsection{Energy partitioning and kinetic energy correction}

\begin{figure}
    \centering
    \includegraphics[width=1\linewidth]{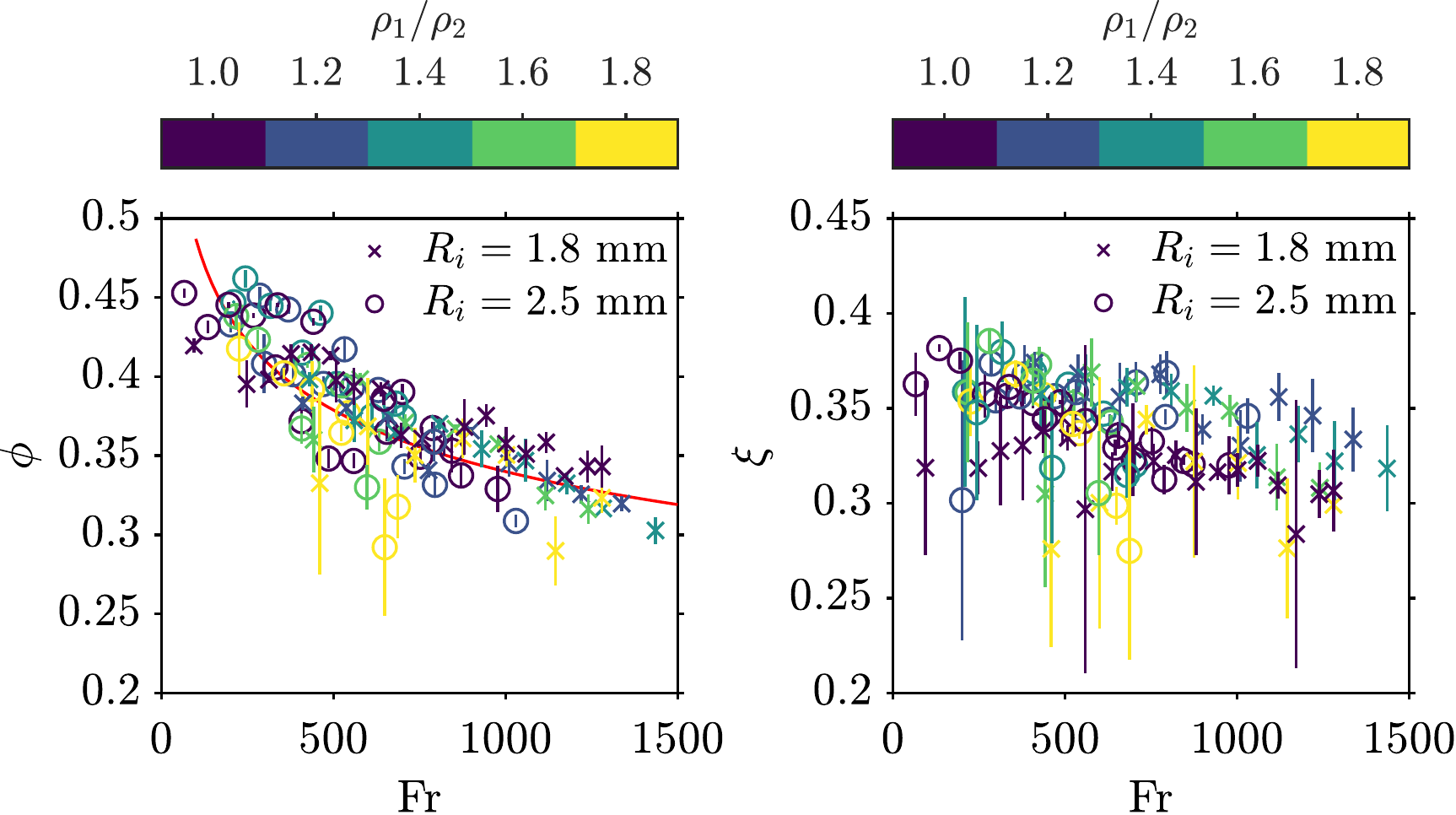}
    \caption{Energy partitioning parameter $\phi$ (a), and crater kinetic energy correction parameter $\xi$ (b), as a function of the Froude number $Fr$. The solid line gives the best-fit power-law scaling (equation \ref{eq:phi_scaling}). Colours scale as the density ratio $\rho_1/\rho_2$. Circles and crosses correspond respectively to large and small drop size series.}
    \label{fig:crater_size_param_01}
\end{figure}

Figure \ref{fig:crater_size_param_01}a shows the correction parameter $\phi$ as a function of the Froude number.
Since the energy is partitioned between the impacting drop and the target, and that several energy sinks such as crown energy and viscous dissipation are neglected in the model, $\phi$ is expected to be smaller than unity.
In our experiments, the energy partitioning coefficient is indeed $\phi=0.38\pm0.04$, in agreement with previous works where experimental data are fitted using a partitioning coefficient in the range $0.2-0.6$, depending on the Froude number \citep{engel_1966,olevson_1969,leng_2001}.

Furthermore, the coefficient $\phi$ is found to be a decreasing function of $Fr$, which scales as
\begin{equation}
    \phi=Fr^{-0.156\pm0.001},
    \label{eq:phi_scaling}
\end{equation}
and is relatively independent of the density ratio and the drop size.
This implies that as the impactor inertia increases, the fraction of kinetic energy delivered to the target decreases. This may be explained by a change in the energy balance between the crater energy and the crown energy \citep{olevson_1969}. As the impactor inertia increases, the relative importance of the surface energy of the crater and the crown decreases, while the potential energy of the crater and the kinetic energy of the crown increases, resulting in a global increase of the crown energy to the expense of the crater. According to \citet{olevson_1969}, the energy within the crown increases with $Fr$ faster than the energy within the crater, which would imply that $\phi$ is a decreasing function of $Fr$.
The drop deformation upon impact may also increase with impactor inertia, and with it the energy required for this deformation, decreasing to this extent the energy delivered to the pool.

Figure \ref{fig:crater_size_param_01}b shows the kinetic energy correction parameter $\xi$, as a function of the Froude number. It accounts for the difference between the deliberately simplified velocity potential used in the model \ref{eq:potential} and the true flow. Since the crater boundary is not hemispherical and the crown is necessarily generated by a tangential velocity field, the true velocity potential in not purely radial, leading to a decrease of the kinetic energy of the flow for a given crater opening velocity \citep{engel_1967,bisighini_2010}. $\xi$ is very likely a function of time, but it is here assumed to be constant.
In our experiments, the kinetic energy correction parameter is smaller than unity with $\xi=0.34\pm0.03$. This means that the velocity model overestimates the crater kinetic energy in the energy balance, as expected. We do not observe any resolvable trend between $\xi$, $Fr$ and $\rho_1/\rho_2$. 

\subsection{Maximum crater radius and opening timescale}
\label{sec:crater_size_maximum}

Assuming that the available impact kinetic energy is fully converted into crater potential energy, \textit{i.e.} neglecting the crater surface energy and the kinetic energy terms, equation \ref{eq:energy_model_adim} gives a scaling for the normalised maximum crater radius
\begin{equation}
    R_{max}^*=\left(\frac{8}{3}\right)^{1/4}\phi^{1/4} Fr^{*^{1/4}}.
    \label{eq:energy_model_Rmax}
\end{equation}

\begin{figure}
    \centering
    \includegraphics[width=1\linewidth]{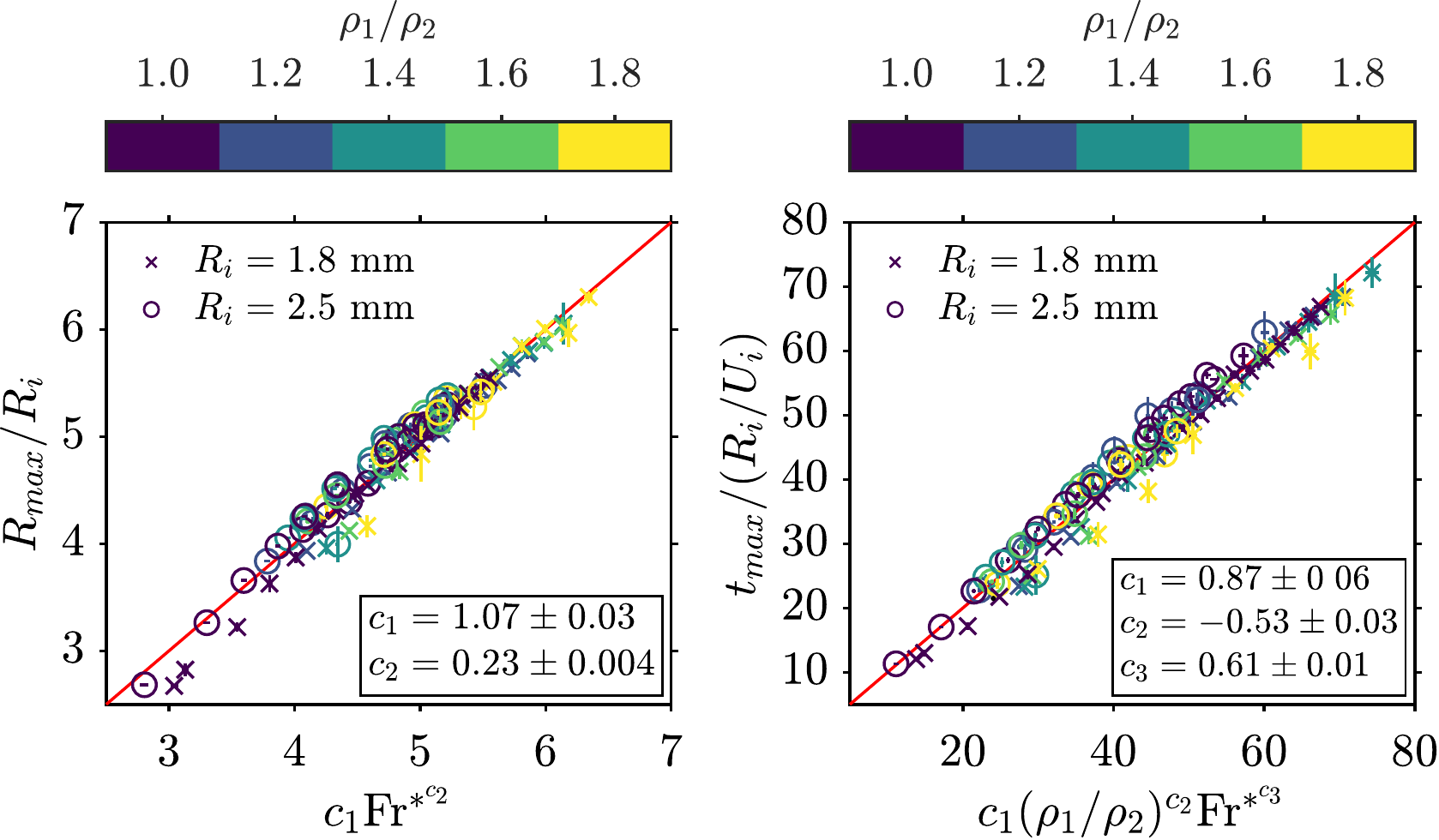}
    \caption{(a) Experimental maximum crater radius $R_{max}$, normalised by the drop radius $R_i$, as a function of the least-squares best-fit power-law scaling, using the modified Froude number $Fr^*$. (b) Experimental crater opening timescale $t_{max}$, normalised by the drop free-fall time $R_i/U_i$, as a function of the least-squares best-fit power-law scaling, using the density ratio $\rho_1/\rho_2$ and the modified Froude number $Fr^*$.
    Colours scale as the density ratio $\rho_1/\rho_2$. Circles and crosses correspond respectively to large and small drop size series.}
    \label{fig:crater_size_max_01}
\end{figure}

Figure \ref{fig:crater_size_max_01}a shows the normalised maximum crater size in our experiments as a function of a least-square best-fit power law scaling in the form $c_1Fr^{*^{c_2}}$.
The exponent $c_2=0.23\pm0.004$ for $Fr^*$ is close to the theoretical $1/4$ prediction of equation \ref{eq:energy_model_Rmax}, and is in agreement with previous works on liquids \citep{prosperetti_1993,leng_2001,bisighini_2010} and granular materials \citep{walsh_2003,takita_2013}.
The prefactor $c_1=1.07\pm0.03$ is close to the value predicted by the model (equation \ref{eq:energy_model_Rmax}). Since $\phi=0.38\pm0.04$ in our experiments, the predicted model prefactor is indeed equal to $1.0\pm0.03$. The prefactor $c_1$ is also consistent with those obtained in previous works (\textit{e.g.} $c_1=1.1$ in \citet{leng_2001}).

We now turn to estimating the crater opening time, defined as the time $t_{max}^*$ at which the maximum crater size is reached. Rearranging equation \ref{eq:energy_model_adim} and integrating between $t=0$ and $t=t_{max}^*$ gives
\begin{equation}
    t_{max}^*= \left(\frac{\rho_1}{\rho_2}\right)^{-1/2} \xi^{1/2} \int_0^{R_{max}^*} \frac{R^{3/2}}{\left(\frac{2}{3}\phi -\frac{1}{4}\frac{R^4}{Fr^*} - \frac{R^2}{Fr^* Bo}\right)^{1/2} } \mathrm{d}R.
\end{equation}
Writing $\tilde{R}=R/R_{max}^*$ with $R_{max}^*$ given by equation \ref{eq:energy_model_Rmax},
\begin{equation}
    t_{max}^*= 2 \left(\frac{8}{3}\right)^{1/8} \left( \frac{\rho_1}{\rho_2} \right)^{-1/2} \phi^{1/8} \xi^{1/2} Fr^{*^{5/8}} \int_0^1 \frac{ \tilde{R}^{3/2} }{ \left(1 - \tilde{R}^4 - \frac{\sqrt{6}}{\sqrt{Fr^* \phi} Bo} \tilde{R}^2\right)^{1/2} } \mathrm{d}\tilde{R}.
\end{equation}
A first estimate of $t_{max}^*$ in the large $Fr^*$ limit can be obtained by neglecting the potential and surface energy terms in equation \ref{eq:energy_model_adim} \citep{leng_2001},
\begin{equation}
    t_{max}^*=\frac{4}{5}\left(\frac{8}{3}\right)^{1/8}\left(\frac{\rho_1}{\rho_2}\right)^{-1/2} \phi^{1/8} \xi^{1/2} Fr^{*^{5/8}}.
    \label{eq:energy_model_tmax}
\end{equation}

Figure \ref{fig:crater_size_max_01}b shows the normalised opening time in our experiments as a function of a least-square best-fit power law scaling in the form $c_1(\rho_1/\rho_2)^{c_2}Fr^{*^{c_3}}$.
The exponent $c_2=-0.53\pm0.03$ for $\rho_1/\rho_2$ agrees with the theoretical $-1/2$ prediction of equation \ref{eq:energy_model_tmax}.
The exponent $c_3=0.61\pm0.01$ for $Fr^*$ is also close to the  $5/8$ prediction of equation \ref{eq:energy_model_tmax}, and agrees with previous works \citep{leng_2001,bisighini_2010}.
The prefactor $c_1=0.87\pm0.06$ is close to the value predicted by equation \ref{eq:energy_model_tmax}, albeit somewhat larger. Since $\phi=0.38\pm0.04$ and $\xi=0.34\pm0.03$ in our experiments, the model prefactor is indeed equal to $0.47\pm0.02$. The prefactor $c_1$ is also somewhat larger than the prefactors obtained in previous works (\textit{e.g.} $c_1=0.59$ in \citet{leng_2001}).

\begin{figure}
    \centering
    \includegraphics[width=1\linewidth]{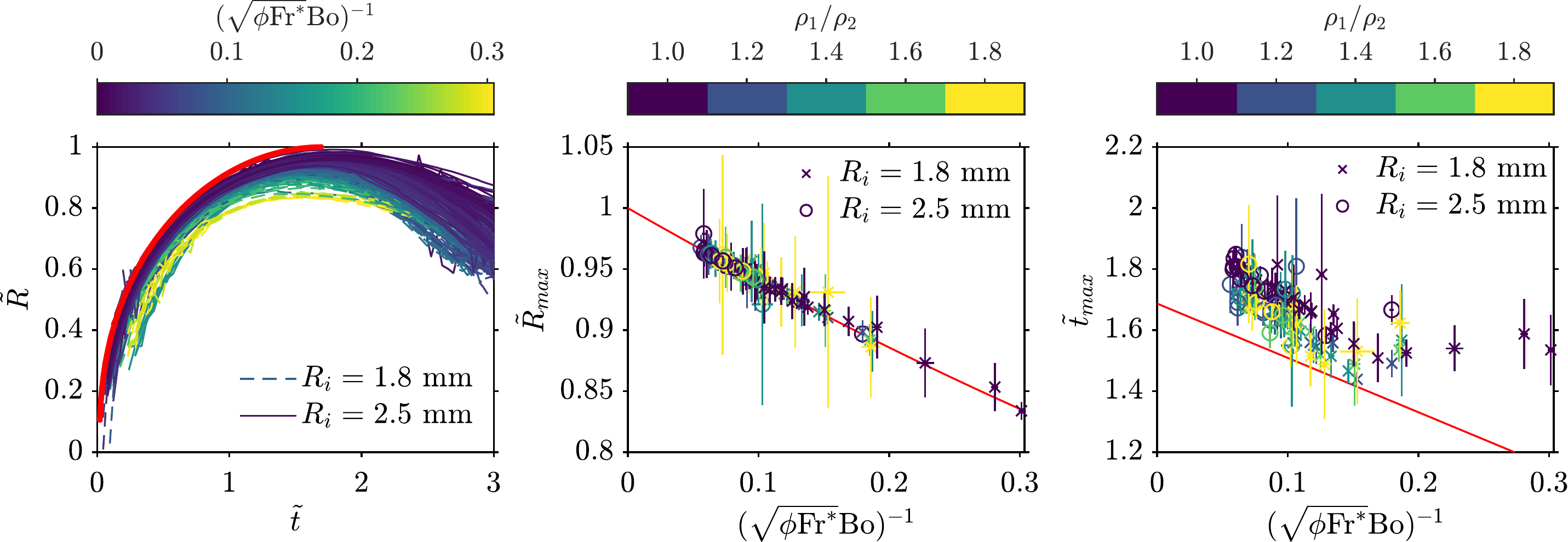}
    \caption{(a) Normalised mean crater radius $\tilde{R}=R/R_{max}^*$, as a function of the normalised time $\tilde{t}=t/t_{max}^*$. The thick solid line gives the master curve of equation \ref{eq:energy_model_renorm} if $1/(\sqrt{\phi Fr^{*}}Bo)=0$. Colours scale as $1/(\sqrt{\phi Fr^{*}}Bo)$ (equation \ref{eq:energy_model_renorm}). Solid and dashed lines correspond respectively to large and small drop size series. (b) Normalised maximum crater radius $\tilde{R}_{max}=R_{max}/R_{max}^*$, as a function of $1/(\sqrt{\phi Fr^{*}}Bo)$. The solid line corresponds to the surface tension correction of equation \ref{eq:energy_model_renorm_Rmax}. Colours scale as the density ratio $\rho_1/\rho_2$. Circles and crosses correspond respectively to large and small drop size series. (c) Normalised crater opening timescale $\tilde{t}=t/t_{max}$, normalised by the predicted crater opening timescale $t_{max}^*$, as a function of $1/(\sqrt{\phi Fr^{*}}Bo)$. The solid line corresponds to the surface tension correction of equation \ref{eq:energy_model_renorm_tmax}. Colours scale as the density ratio $\rho_1/\rho_2$. Circles and crosses correspond respectively to large and small drop size series.}
    \label{fig:crater_size_norm_01}
\end{figure}

By renormalising the crater radius and time as $\tilde{R}=R/R_{max}^*$ and $\tilde{t}=t/t_{max}^*$, respectively, equation \ref{eq:energy_model_adim} gives
\begin{equation}
    \tilde{R}^4+\frac{\sqrt{6}}{\sqrt{Fr^*\phi}Bo}\tilde{R}^2+\frac{25}{4}\tilde{R}^3\tilde{\dot{R}}^2=1,
    \label{eq:energy_model_renorm}
\end{equation}
which depends only on the dimensionless parameter $\sqrt{Fr^*\phi}Bo$. This parameter brings in the effect of surface tension on the cratering dynamics. This is a second order effect compared to the scalings of equations \ref{eq:energy_model_Rmax} and \ref{eq:energy_model_tmax}.
Figure \ref{fig:crater_size_norm_01}a shows the time evolution of the crater radius normalised that way. The energy model (equation \ref{eq:energy_model_adim}) applies only during the opening of the crater, \textit{i.e.} when $R<R_{max}^*$. Hence, experimental data collapse only when $t<t_{max}^*$, with a residual dependency on $\sqrt{Fr^*\phi}Bo$.

At $\tilde{R}=\tilde{R}_{max}$, the crater opening velocity is zero, which using equation \ref{eq:energy_model_renorm} gives a corrected scaling law accounting for surface tension effects on the maximum crater radius
\begin{equation}
    \tilde{R}_{max}= \left[\sqrt{1+\frac{3}{2}(\sqrt{Fr^*\phi}Bo)^{-2}}-\frac{\sqrt{6}}{2}(\sqrt{Fr^*\phi}Bo)^{-1}\right]^{1/2}.
    \label{eq:energy_model_renorm_Rmax}
\end{equation}
By integrating equation \ref{eq:energy_model_renorm}, a corrected estimate accounting for surface tension is obtained for the opening timescale
\begin{equation}
    \tilde{t}_{max}=\frac{5}{2}\int_0^{\tilde{R}_{max}}\tilde{R}^{3/2}\left(1-\tilde{R}^4-\frac{\sqrt{6}}{\sqrt{Fr^*\phi}Bo}\tilde{R}^2\right)^{-1/2}\mathrm{d}\tilde{R}.
    \label{eq:energy_model_renorm_tmax_integral}
\end{equation}
Substituting $\tilde{R}$ by $\tilde{R}/\tilde{R}_{max}$ in equation \ref{eq:energy_model_renorm_tmax_integral}, where $\tilde{R}_{max} \sim 1-\frac{\sqrt{6}}{4}(\sqrt{Fr^*\phi}Bo)^{-1}$ is the first-order development of equation \ref{eq:energy_model_renorm_Rmax}, and developing a first-order approximation of $\tilde{t}_{max}$ as a function of $(\sqrt{Fr^*\phi}Bo)^{-1}$, equation \ref{eq:energy_model_renorm_tmax_integral} gives
\begin{equation}
    \tilde{t}_{max}=\frac{5}{8}\mathrm{B}\left(\frac{1}{2},\frac{5}{8}\right)-\frac{5\sqrt{6}}{64}\mathrm{B}\left(\frac{1}{2},\frac{1}{8}\right)(\sqrt{Fr^*\phi}Bo)^{-1},
    \label{eq:energy_model_renorm_tmax}
\end{equation}
where $\mathrm{B}$ is the beta function, $\frac{5}{8}\mathrm{B}\left(\frac{1}{2},\frac{5}{8}\right) \approx 1.687$, and $\frac{5\sqrt{6}}{64}\mathrm{B}\left(\frac{1}{2},\frac{1}{8}\right) \approx 1.781$.
If surface tension is neglected, \textit{i.e.} $Bo \to +\infty$, $\tilde{t}_{max}=\frac{5}{8}\mathrm{B}\left(\frac{1}{2},\frac{5}{8}\right)$ is an exact solution of equation \ref{eq:energy_model_renorm_tmax_integral}.

Figure \ref{fig:crater_size_norm_01}b shows the maximum normalised crater radius $\tilde{R}_{max}$, as a function of $(\sqrt{Fr^*\phi}Bo)^{-1}$. It corresponds to the ratio between experimental data and the scaling law without surface tension (equation \ref{eq:energy_model_Rmax}).
As expected, the scaling overestimates the experimental maximum crater radius because it neglects surface energy, and several energy sinks related to the crown formation. This overestimate decreases with $(\sqrt{Fr^*\phi}Bo)^{-1}$, \textit{i.e.} when surface tension effects become negligible in comparison with impactor inertia and gravity forces.
The difference between experimental data and the scaling law without surface tension is properly corrected by equation \ref{eq:energy_model_renorm_Rmax}, using the surface tension term.

Figure \ref{fig:crater_size_norm_01}c shows the normalised crater opening timescale $\tilde{t}_{max}$, as a function of $(\sqrt{Fr^*\phi}Bo)^{-1}$. It corresponds to the ratio between experimental data and the the scaling law without surface tension (equation \ref{eq:energy_model_tmax}). The scaling underestimates the experimental opening timescale. When surface energy and crown formation are neglected, the crater is indeed expected to open faster, \textit{i.e.} on a reduced timescale. The difference between experimental data and the scaling law without surface tension is partly corrected by equation \ref{eq:energy_model_renorm_tmax}, giving a reasonable trend and relative errors under 20\%.

\section{Evolution of the Rayleigh-Taylor instability}
\label{sec:mixing_layer}

Experiments in the LIF configuration provide the time evolution of the mixing layer. Using the energy conservation model, a model for the mixing layer thickness evolution is derived and compared with experiments. A linear stability analysis model is also used to obtain theoretical instability wavelength at early times.

\subsection{Mixing layer thickness}

\subsubsection{Mixing model}

In addition to the energy conservation model assumptions, we assume the mixing layer to be homogeneous with a constant thickness around the crater boundary. 

We first consider a situation where negligible mixing occurs between the drop liquid and its surrounding. As the crater radius increases, the drop liquid spreads over an increasingly large surface area, decreasing in turn its mean thickness $h$. Denoting by $\bar{u}(r,t)$ the laterally averaged velocity field associated with the opening of the crater, the time derivative of $h$ is then given by
\begin{equation}
    \dot{h}=\bar{u}(R+h)-\bar{u}(R).
    \label{eq:mixing_conservation}
\end{equation}
Since $\bar{u}=\dot{R}(R/r)^2$ corresponds to the radial potential flow of equation \ref{eq:velocity}, this gives
\begin{equation}
    \dot{h}=\dot{R}\left[\frac{R^2}{(R+h)^2}-1\right],
    \label{eq:mixing_spreading_term}
\end{equation}
where the right-hand side is referred to as a spreading term. This equation corresponds to the mass conservation of the layer.

\begin{figure}
    \centering
    \includegraphics[width=\linewidth]{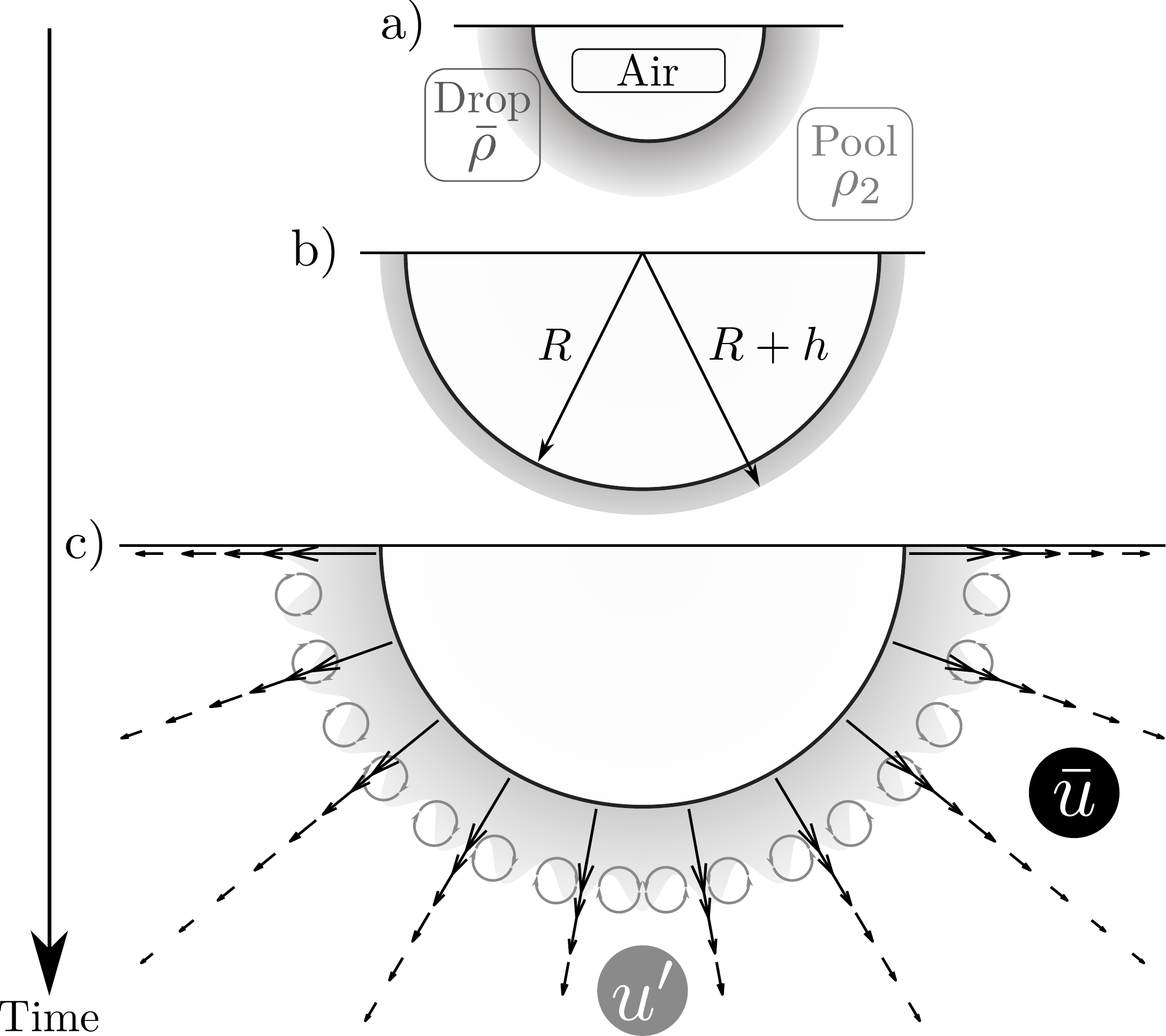}
    \caption{Mixing layer evolution.
    After having spread quickly on the crater boundary to become a thick layer (a), the liquid layer of the drop gradually gets thinner as the crater grows (b). At some point, crater expansion becomes weak enough, allowing for the Rayleigh-Taylor instability to develop (c).
    The velocity field is decomposed into a velocity component $\bar{u}$ produced by crater opening, and velocity fluctuations $u'$ produced by the Rayleigh-Taylor instability.}
    \label{fig:model_sketch}
\end{figure}

We now consider a simple model of RT induced mixing (figure \ref{fig:model_sketch}), assuming that the instability produces velocity fluctuations in the mixing layer. We thus define a scalar mixing term $u'(r,t)$, which corresponds to a measure of these velocity fluctuations, and represents the mixing intensity. This mixing term corresponds to an inward flux of ambient liquid toward the mixing layer.
% and conversely to an outward flux of dense material toward the ambient fluid, both effects contributing to increase the thickness of the mixing layer.
We then assume $\dot{h}$ to be the sum of the spreading and mixing terms (figure \ref{fig:model_sketch}c), which gives
\begin{equation}
    \dot{h}=\dot{R}\left[\frac{R^2}{(R+h)^2}-1\right]+u'.
    \label{eq:mixing_conservation_adim}
\end{equation}
The velocity $u'$ and length scale $h$ can be seen as the velocity and integral length-scale of a mixing-length turbulent model describing the mixing layer.

The evolution of $u'$ is obtained using a buoyancy-drag model of the mixing layer \citep{dimonte_2000a,oron_2001,zhao_2013}. If we consider that the mixing layer with a density $\bar{\rho}=\rho_2+\Delta\rho$ penetrates into the less dense surrounding liquid with a density $\rho_2$, an equation of motion is
\begin{equation}
    \bar{\rho}\frac{\mathrm{d}u'}{\mathrm{d}t}=\beta\Delta\rho|\ddot{R}|-C\rho_2\frac{u'^2}{h},
    \label{eq:mixing_buoy_drag}
\end{equation}
where $\beta$ and $C$ are the RT buoyancy prefactor and the drag coefficient, respectively.
This equation corresponds to a balance between the fluid inertia on the left-hand side, buoyancy in the first term of the right-hand side, and inertial drag in the second term of the right-hand side.
The acceleration of the crater boundary $\ddot{R}$ being significantly larger than $g$, Earth's gravity is neglected in the buoyancy term. 

Using mass conservation in the homogeneous mixing layer, the dimensionless density evolution is
\begin{equation}
    \frac{\Delta\rho}{\bar{\rho}}=\frac{1}{1+\frac{3}{2}R^2h\frac{\rho_2}{\Delta\rho_0}},
    \label{eq:density_evolution}
\end{equation}
where $\Delta\rho_0$ is the initial density difference, \textit{i.e.} between the impacting drop and the pool.
Replacing the density evolution (equation \ref{eq:density_evolution}) in the equation of motion eventually gives
\begin{equation}
    \frac{\mathrm{d}u'}{\mathrm{d}t}=\beta\frac{|\ddot{R}|}{1+\frac{3}{2}R^2h\frac{\rho_2}{\Delta\rho_0}}-C\frac{1}{1+\frac{2}{3}\frac{1}{R^2h}\frac{\Delta\rho_0}{\rho_2}}\frac{u'^2}{h}.
    \label{eq:mixing_buoy_drag_adim}
\end{equation}

Together with the crater radius evolution (equation \ref{eq:energy_model_adim}), equations \ref{eq:mixing_conservation_adim} and \ref{eq:mixing_buoy_drag_adim} are coupled ordinary differential equations.
This initial value problem is solved numerically using $R(1)=1$, $h(1)=2/3$ and $u'(1)=0$ as initial conditions. For each experiment, the experimentally measured crater radius $R(t)$ and mixing layer thickness $h(t)$ are used to determine the best value for the fitting parameters, which are the energy partitioning coefficient $\phi$, the kinetic energy correction coefficient $\xi$, the buoyancy prefactor $\beta$, and the drag coefficient $C$.

\subsubsection{Time evolution}

\begin{figure}
    \centering
    \includegraphics[width=1\linewidth]{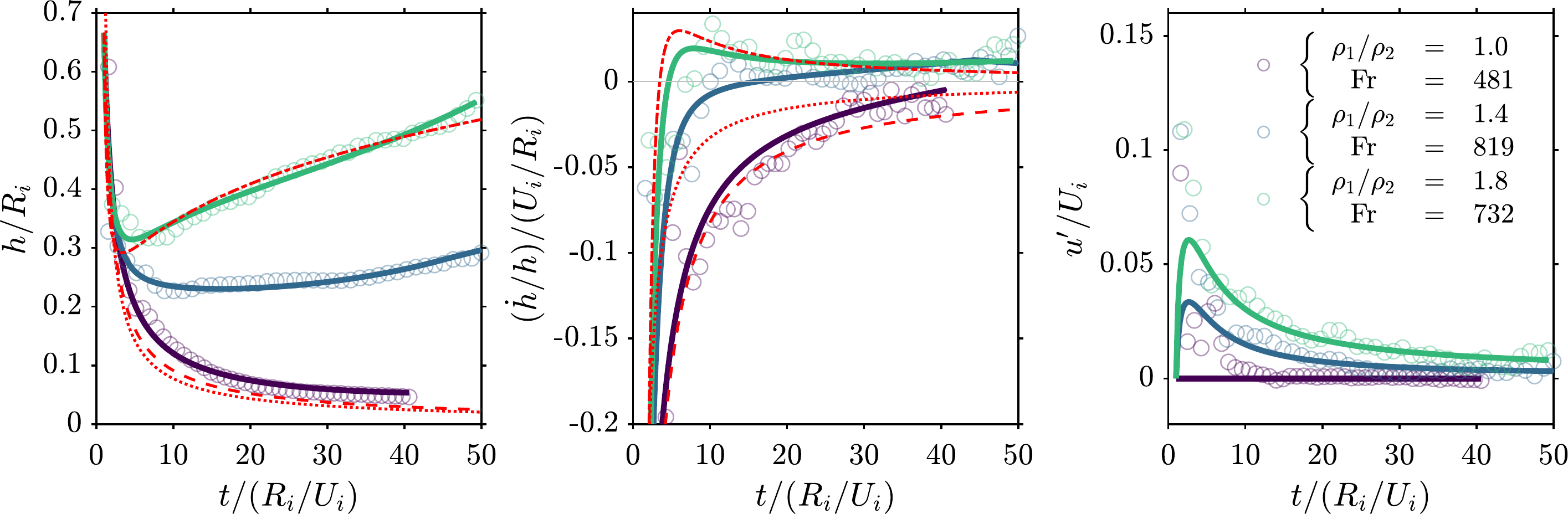}
    \caption{Time evolution, normalised by the drop free-fall time $R_i/U_i$, of the mixing layer thickness $h$ normalised by the drop radius $R_i$ (a), the mixing layer growth rate $\dot{h}/h$ normalised by the drop free-fall rate $U_i/R_i$ (b), and the estimated inward flux density due to mixing $u'$ (from equation \ref{eq:mixing_conservation_adim}) normalised by the impact velocity $U_i$ (c), for three impact parameters. Circles, solid lines, dashed lines, dotted lines, and dash-dotted lines correspond respectively to experimental data, fitted mixing model (equations \ref{eq:energy_model_adim}, \ref{eq:mixing_conservation_adim} and \ref{eq:mixing_buoy_drag_adim}), complete early-time power-law analytical solution (equation \ref{eq:mixing_model_solution}), approximate early-time power-law analytical solution (equation \ref{eq:mixing_model_solution_approx}), and late-time power-law analytical solution (equation \ref{eq:h_power_mixing} with $h_0=2/3$, $\dot{h}_0=-1.25$, and $C=0.70$). Analytical solutions are calculated for the $\rho_1/\rho_2=1.8$ and $Fr=732$ experiment.}
    \label{fig:h_time_01}
\end{figure}

Figure \ref{fig:h_time_01} compares the fitted mixing layer evolution model with experimental data, in three reference cases, without density difference and with two density ratios of the impacting drop and the pool. In figure \ref{fig:h_time_01}c, $u'$ is estimated from equation \ref{eq:mixing_conservation_adim} based on experimental measurements of $R$ and $h$.
The fitted time evolution of the mixing layer model is most of the time in agreement with experimental data. For dimensionless time typically smaller than 10, \textit{i.e.} at the very beginning of the crater expansion, the model often overestimates the mean mixing layer thickness. It may be explained by model assumptions being poorly verified, such as a nearly hemispherical crater and an homogeneous mixing layer (\textit{e.g.} figure \ref{fig:sequence_01}).
In the same way, in figure \ref{fig:h_time_01}c, $u'$ is overestimated for dimensionless times typically smaller than 10, for experiments with and without density difference. It means that the measured mixing layer growth rate $\dot{h}$ is larger than the predicted velocity related to the geometrical evolution of the crater (equation \ref{eq:mixing_conservation}).
This overestimate is explained by the initial accumulation of the liquid of the drop at the crater floor. The liquid of the drop flows on the crater sides, producing a mixing layer growth rate comparatively larger than in the model with an homogeneous thickness. The overestimate indeed decreases on a timescale corresponding to the drop spread on the crater sides.

\subsubsection{Buoyancy prefactor and drag coefficient}

\begin{figure}
    \centering
    \includegraphics[width=1\linewidth]{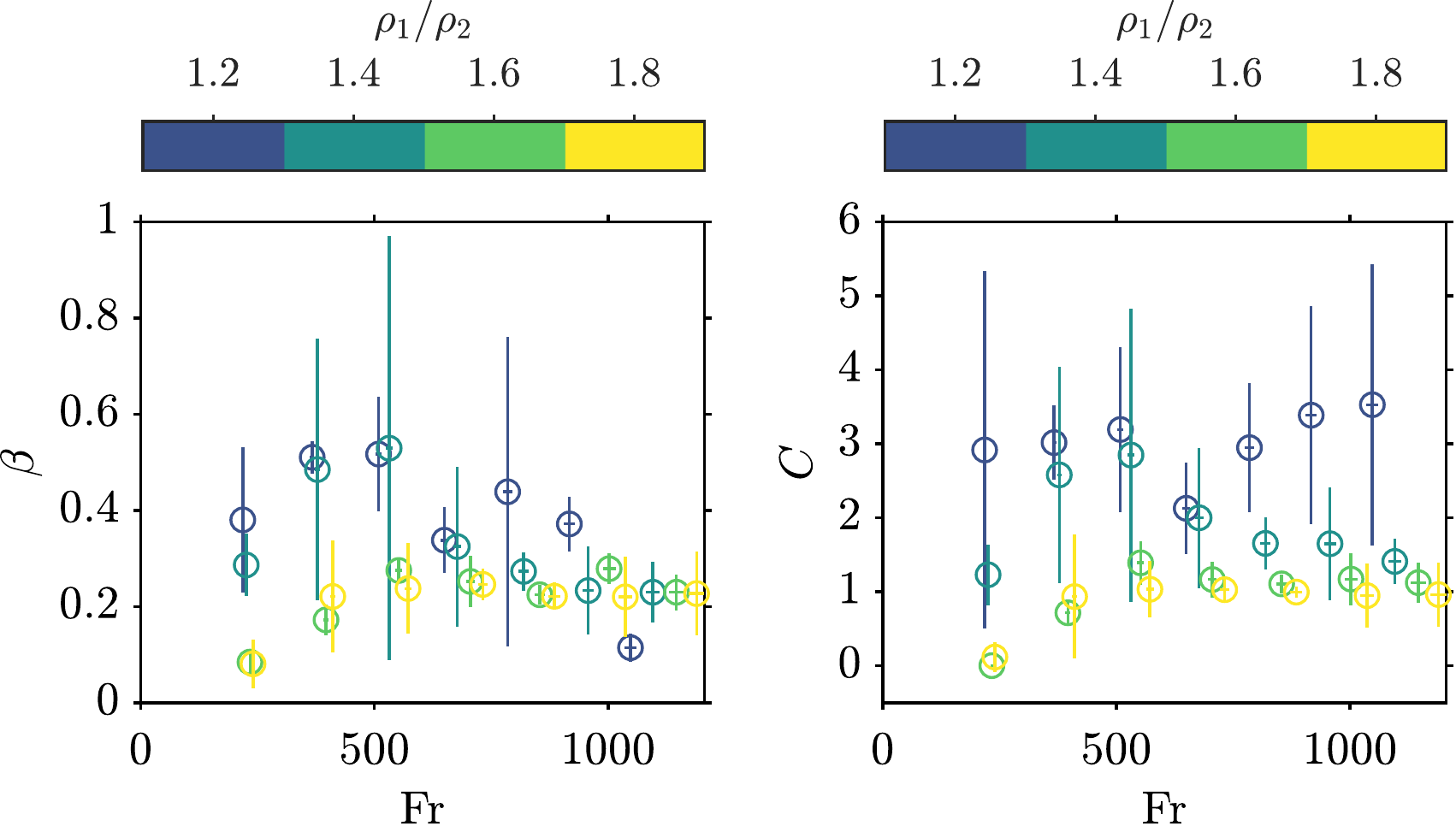}
    \caption{Buoyancy prefactor $\beta$ (a), and drag coefficient $C$ (b), as a function of the Froude number $Fr$. Colours scale as the density ratio $\rho_1/\rho_2$.}
    \label{fig:buoy_drag_01}
\end{figure}

Figure \ref{fig:buoy_drag_01} shows the fitted buoyancy prefactor $\beta$ and drag coefficient $C$ for each experiment. Our results are compared with the plane layer experiments of \cite{dimonte_2000a}, who found a good agreement between their experiments and buoyancy-drag model with $\beta=1$ and $C=2.5\pm0.6$.

In our experiments, the mean value of the buoyancy prefactor is $\beta=0.3\pm0.1$, \textit{i.e.} smaller than unity. This difference in the buoyancy prefactor may be interpreted as a consequence of the spherical interface, or the limited thickness of the dense layer. Given error bars, we find that $\beta$ may be independent of the density ratio, but may increase with the Froude number at large density ratios ($\rho_1/\rho_2=\{1.6,1.8\}$), until it reaches $\beta\simeq0.25$ at $Fr=500$.

The mean value of the drag coefficient is $C=2\pm1$. This value is in agreement with the value $C=2.5\pm0.6$  obtained for constant, variable and impulsive accelerations history \citep{dimonte_2000a}. Given error bars, we find that $C$ may decrease when the density ratio increases, the mean value of the drag coefficient at $\rho_1/\rho_2=1.2$ and $\rho_1/\rho_2=1.8$ being respectively $C=3.0\pm0.5$ and $C=0.9\pm0.3$. As for the buoyancy prefactor, $C$ may increase with the Froude number at large density ratios ($\rho_1/\rho_2=\{1.6,1.8\}$), until it reaches $C\simeq1$ at $Fr=500$. 

In order to interpret and compare the values of $\beta$ and $C$, two simplified acceleration history, different from the actual acceleration of the crater, are now considered: a constant acceleration and an impulsive acceleration.

\begin{figure}
    \centering
    \includegraphics[width=1\linewidth]{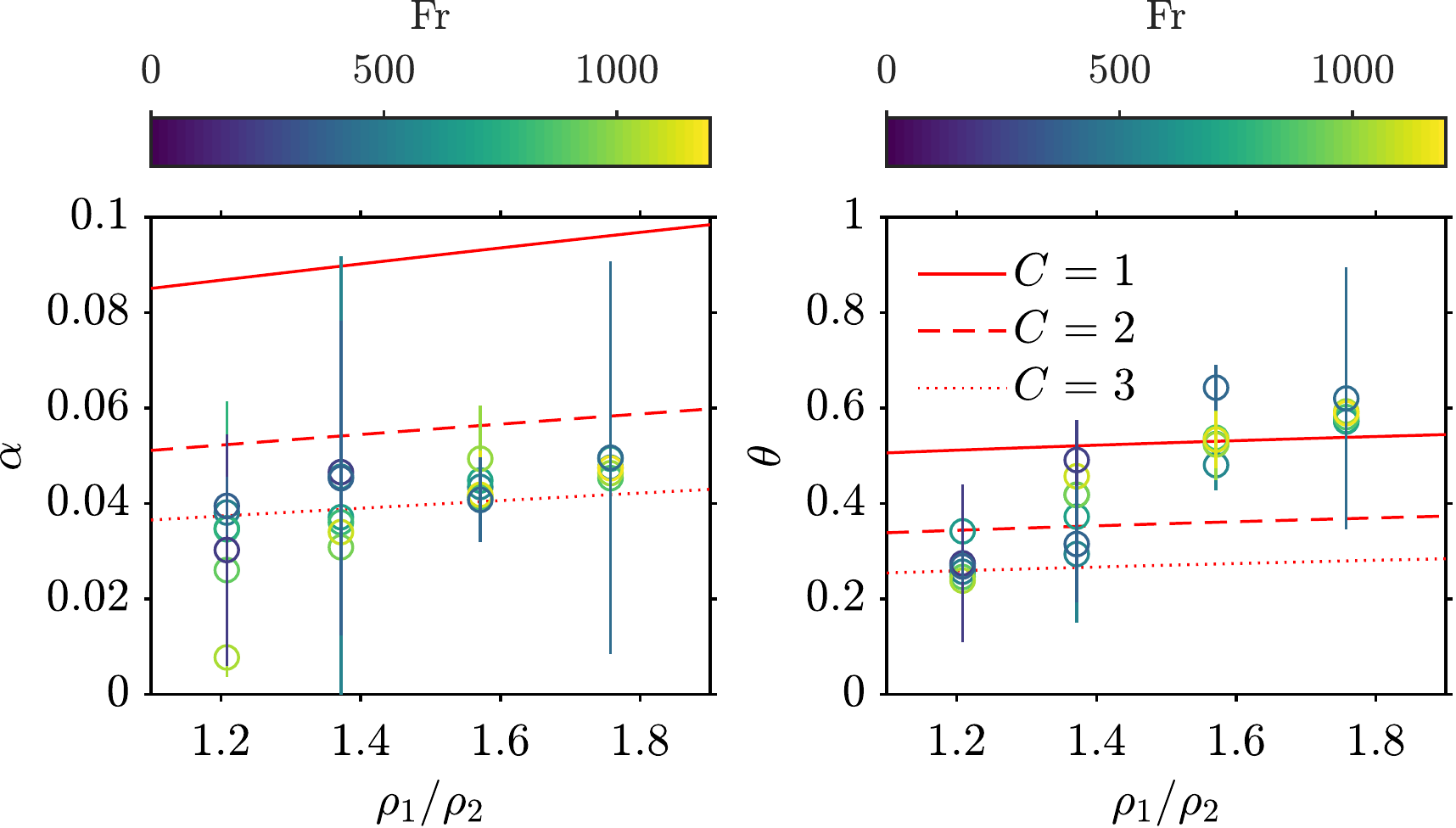}
    \caption{Constant acceleration prefactor $\alpha$ (a), and impulsive acceleration exponent $\theta$ (b), as a function of the density ratio $\rho_1/\rho_2$. Colours scale as the Froude number $Fr$. Solid lines, dashed lines, and dotted lines correspond to the homogeneous buoyancy-drag model of \citet{dimonte_2000a} for $C=1$, $C=2$, and $C=3$, respectively.}
    \label{fig:alpha_theta_01}
\end{figure}

In the case of a constant acceleration $|\ddot{R}|$, the solution to the buoyancy-drag equation \ref{eq:mixing_buoy_drag} in which geometrical effects are neglected is $h=\alpha(\Delta\rho/\bar{\rho})|\ddot{R}|t^2$ \citep[\textit{e.g.}][]{dimonte_2000a}, where $\alpha$ is a prefactor depending on several parameters such as fluid miscibility, and the fluid being penetrated, \textit{i.e.} light into heavy or heavy into light. Since $u'=\dot{h}$, and assuming that $\bar{\rho}=(\rho_1+\rho_2)/2$, the prefactor $\alpha$ is
\begin{equation}
    \alpha=\frac{\beta}{2+8C\frac{\rho_2}{\rho_1+\rho_2}}.
\end{equation}
Figure \ref{fig:alpha_theta_01}a shows $\alpha$, calculated for each experiment, and compares the results to the homogeneous buoyancy-drag model of \citet{dimonte_2000a} calculated for $C=1$, $C=2$, and $C=3$.
The mean value $\alpha=0.04\pm0.01$ is somewhat smaller than values (0.05-0.07) measured at the same density ratio between immiscible fluids, with a constant acceleration \citep{dimonte_2000}.
Within error bars, $\alpha$ may also increase with the density ratio. However, the homogeneous model of \citet{dimonte_2000a} overestimates the observed values of $\alpha$, in particular for large density ratio. For example, at $\rho_1/\rho_2=1.8$ the drag coefficient is approximately $C=1$ (figure \ref{fig:buoy_drag_01}b), which leads to an overestimate of $\alpha$ by a factor 2.
This may be a consequence of the variable acceleration, but also of the spherical interface, miscibility, and the limited thickness of the dense layer.

In the case of an impulsive acceleration, the buoyancy term in equation \ref{eq:mixing_buoy_drag} is negligible since $|\ddot{R}|=0$. Neglecting geometrical effects, \textit{i.e.} $u'=\dot{h}$, and assuming that $\bar{\rho}=(\rho_1+\rho_2)/2$, the solution is then given by $h=h_0\tau^\theta$, where $\tau=u'_0t/\theta h_0+1$, and $h_0$ and $u'_0$ are initial values \citep[\textit{e.g.}][]{dimonte_2000a}. The exponent is then
\begin{equation}
    \theta=\frac{1}{1+2C\frac{\rho_2}{\rho_1+\rho_2}}.
\end{equation}
Figure \ref{fig:alpha_theta_01}b shows $\theta$, calculated for each experiments, and compares the results to the buoyancy-drag model of \citet{dimonte_2000a}. The mean value $\theta=0.4\pm0.1$ is close to values (0.2-0.3) measured at the same density ratio between immiscible fluids, with an impulsive acceleration \citep{dimonte_2000}.
Within error bars, $\theta$ increases with the density ratio, consistently with the homogeneous model of \citet{dimonte_2000a} estimated at compatible values of the drag coefficient $C$ (figure \ref{fig:buoy_drag_01}).
Since the acceleration of the crater is approximately a $t^{-8/5}$ power-law, the acceleration is relatively close to be impulsive, explaining the good agreement between our experiments and the impulsive acceleration model.

\subsubsection{Geometrical stage, mixing stage, and transition timescale}

\begin{figure}
    \centering
    \includegraphics[width=1\linewidth]{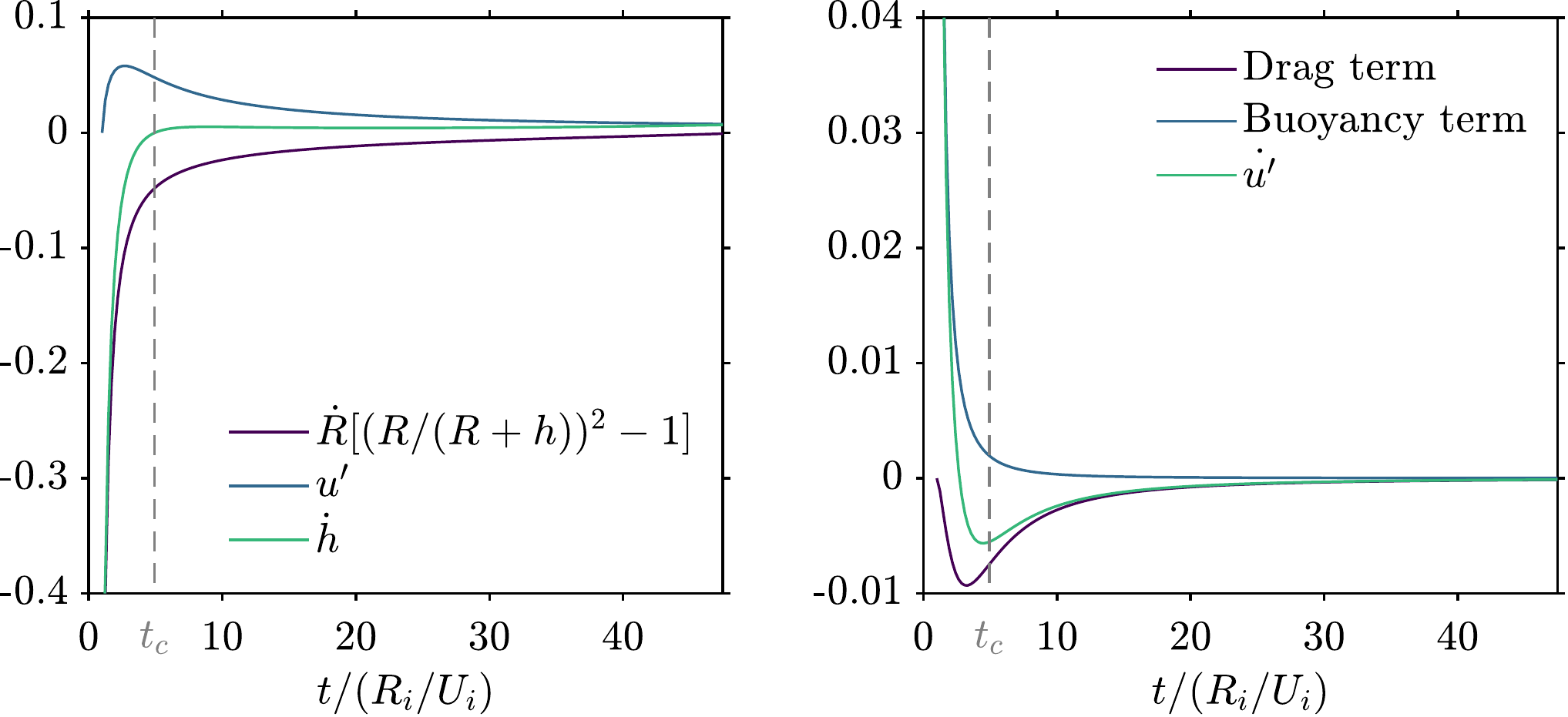}
    \caption{(a) Spreading term (first term on the right-hand side), mixing term (second term on the right-hand side), and $\dot{h}$ (left-hand side) in equation \ref{eq:mixing_conservation_adim}, as a function of time, normalised by the drop free-fall time $R_i/U_i$.
    (b) Buoyancy term (first term on the right-hand side), drag term (second term on the right-hand side), and $\dot{u'}$ (left-hand side) in equation \ref{eq:mixing_buoy_drag_adim}, as a function of time, normalised by the drop free-fall time $R_i/U_i$. In this experiment, $\rho_1/\rho_2=1.8$ and $Fr=572$.}
    \label{fig:model_terms_01}
\end{figure}

We now focus on the numerical solution of the coupled ordinary differential equations.
Using figure \ref{fig:model_terms_01}, which shows the spreading term (first term on the right-hand side) and the mixing term (second term on the right-hand side) in equation \ref{eq:mixing_conservation_adim}, and the buoyancy term (first term on the right-hand side) and the drag term (second term on the right-hand side) in equation \ref{eq:mixing_buoy_drag_adim}, as a function of time, the mixing layer evolution is decomposed into a geometrical stage and a mixing stage with a transition happening at a time $t_c$.

The first stage ($t<t_c$), referred to as the geometrical stage, corresponds to a negative growth rate of the mixing layer. Its dynamics is controlled by the geometrical evolution of the crater, with the spreading term prevailing over the RT mixing term in equation \ref{eq:mixing_conservation_adim}. Since the crater deceleration is large at early times, the buoyancy term prevails over the drag term in the buoyancy-drag equation \ref{eq:mixing_buoy_drag_adim}. However, this does not influence the evolution of the mixing layer since the spreading term prevails.

In this stage, if the mixing term $u'$ is neglected, the solution of equation \ref{eq:mixing_conservation_adim} is 
\begin{equation}
    h=\left(\frac{98}{27}+R^3\right)^{1/3}-R.
    \label{eq:mixing_model_solution}
\end{equation}
Using the power-law solution of $R$ (equation \ref{eq:energy_model_assum}), an analytical solution is obtained for $h$ in the geometrical phase (figure \ref{fig:h_time_01}, dashed lines).
If in addition $h \ll R$, what might be a reasonable assumption after a few dimensionless time (\textit{e.g.} figure \ref{fig:sequence_01}), equation \ref{eq:mixing_conservation_adim} simplifies as 
\begin{equation}
    \dot{h}=-2\dot{R}\frac{h}{R}.
    \label{eq:mixing_spreading_term_approx}
\end{equation}
The solution corresponds to the conservation of the initial volume of the impactor, a sphere of unit dimensionless radius, which takes the form
\begin{equation}
    h=\frac{2}{3}\frac{1}{R^2}.
    \label{eq:mixing_model_solution_approx}
\end{equation}
Using the power-law solution of equation \ref{eq:energy_model_assum}, equation \ref{eq:mixing_model_solution_approx} also gives a power-law solution for the mixing layer thickness, velocity, and acceleration 
\begin{equation}
    \left\{
    \begin{array}{l}
        h(t)=\frac{2}{3}\left[Q(t-1)+1\right]^{-4/5}\\
        \dot{h}(t)=-\frac{8}{15}Q\left[Q(t-1)+1\right]^{-9/5}\\
        \ddot{h}(t)=\frac{24}{25}Q^2\left[Q(t-1)+1\right]^{-14/5}\\
    \end{array}
    \right.,
    \label{eq:mixing_model_early_thickness}
\end{equation}
and consequently a power-law solution for the mixing layer growth rate
\begin{equation}
    \frac{\dot{h}}{h}(t)=-\frac{4}{5}Q\left[Q(t-1)+1\right]^{-1}.
    \label{eq:mixing_model_early_rate}
\end{equation}
These solutions (figure \ref{fig:h_time_01}, dotted lines) depend on the density ratio $\rho_1/\rho_2$, and on the correction parameters $\phi$ and $\xi$, through $Q$.

\begin{figure}
    \centering
    \includegraphics[width=1\linewidth]{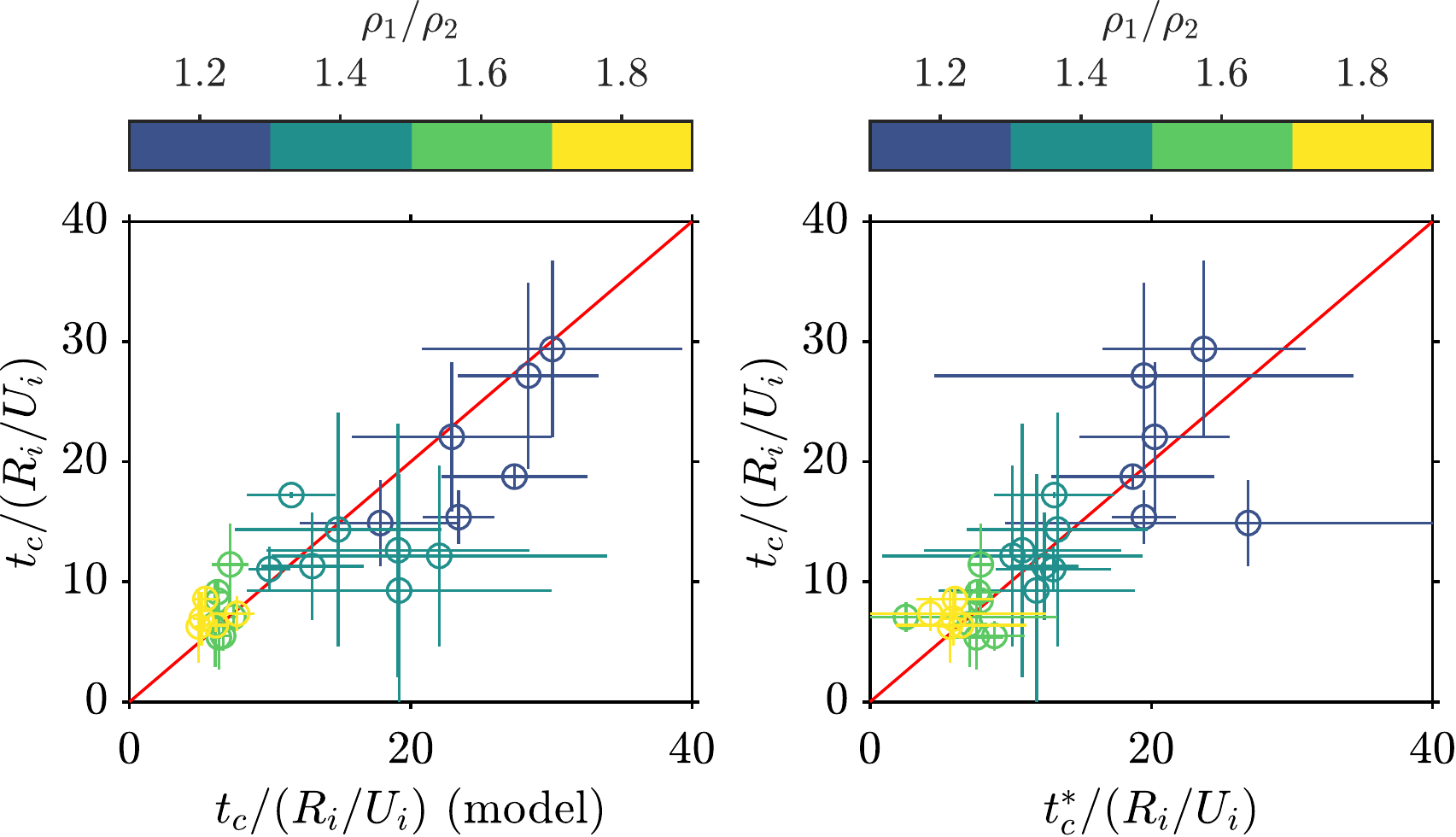}
    \caption{(a) Experimental transition timescale $t_c$, normalised by the drop free-fall time $R_i/U_i$, as a function of the transition time predicted by the mixing model (equations \ref{eq:energy_model_adim}, \ref{eq:mixing_conservation_adim} and \ref{eq:mixing_buoy_drag_adim}), normalised by the drop free-fall time $R_i/U_i$.
    (b) Experimental transition timescale $t_c$, normalised by the drop free-fall time $R_i/U_i$, as a function of the transition time scaling $t_c^*$ (equation \ref{eq:mixing_model_tc}), normalised by the drop free-fall time $R_i/U_i$. Colours scale as the density ratio $\rho_1/\rho_2$.}
    \label{fig:tc_01}
\end{figure}

The transition time $t_c$ between the geometrical stage and the mixing stage can be defined as the time at which geometrical effects are of the same order of magnitude as the mixing produced by the RT instability, \textit{i.e.} when the growth rate changes sign and the mixing layer thickness reaches a local minimum.
The transition time between stages is determined experimentally and compared with the transition time obtained from the numerical model (figure \ref{fig:tc_01}a). Although uncertainties on the transition time are significant due to the extrinsic variability of experiments in the same configuration, the numerical model is rather consistent with experimental data.

An analytical estimate for the transition time $t_c^*$ is also derived.
Equation \ref{eq:mixing_conservation_adim} is simplified using approximation of equation \ref{eq:mixing_spreading_term_approx} for the geometrical term, \textit{i.e.} $h \ll R$, which is supposed to be valid at the transition time.
Equation \ref{eq:mixing_buoy_drag_adim} is also simplified assuming $\frac{2}{3}\frac{\Delta\rho_0}{\rho_2}\frac{1}{R^2h} \ll 1$. These assumptions correspond to $\Delta\rho \ll \rho_2$, which might be a reasonable assumption after at the transition time. Using these assumptions, equations \ref{eq:mixing_conservation_adim} and \ref{eq:mixing_buoy_drag_adim} give
\begin{equation}
    \ddot{h} + 2(2C+1)\frac{\dot{R}}{R}\dot{h} + 2(2C-1)\frac{\dot{R}^2}{R^2}h + C\frac{\dot{h}^2}{h} + 2h\frac{\ddot{R}}{R} - \beta\frac{2}{3}\frac{\Delta\rho_0}{\rho_2}\frac{|\ddot{R}|}{R^2h} = 0.
    \label{eq:mixing_model_tc_equation}
\end{equation}
At the critical transition time $t_c$, geometrical effects are of the same order of magnitude as mixing effects, which means that $\dot{h}(t=t_c)=0$ (equation \ref{eq:mixing_conservation_adim}). Using in addition the power-laws solutions of $R$ (equation \ref{eq:energy_model_assum}) and $h$ (equation \ref{eq:mixing_model_early_thickness}) in the geometrical stage, an estimate of $t_c$ is obtained from equation \ref{eq:mixing_model_tc_equation}
\begin{equation}
    t_c^*=c\left\{1+\frac{1}{Q}\left[\left(\frac{16}{9}\frac{C+1}{\beta\frac{\Delta\rho_0}{\rho_2}}\right)^{5/6}-1\right]\right\},
    \label{eq:mixing_model_tc}
\end{equation}
where $c=1.09\pm0.06$ is a least-squares best-fit prefactor obtained from experimental data (figure \ref{fig:tc_01}b). As expected in figure \ref{fig:h_time_01}, the transition time decreases with the density contrast between the fluid of the impacting drop and the pool. The larger the density contrast, the quicker mixing effects become comparable to geometrical effects.

The second stage ($t>t_c$), referred to as the mixing stage, corresponds to a positive growth rate of the mixing layer. Its dynamics is controlled by a balance between residual geometrical effects and mixing produced by the RT instability. The spreading term and the mixing term are then similar in magnitude. Since the drag term prevails over the buoyancy term because the crater deceleration vanishes, the mixing term is consequently controlled by the drag term. At late times, when the crater is close to reach its maximum, the spreading term vanishes and the mixing layer dynamics is only controlled by the mixing term.

In this stage, equation \ref{eq:mixing_conservation_adim} may be simplified using approximation of equation \ref{eq:mixing_spreading_term_approx} for the geometrical term, which is supposed to be valid in the mixing stage.
Equation \ref{eq:mixing_buoy_drag_adim} is simplified neglecting the buoyancy term and 
assuming $\frac{2}{3}\frac{\Delta\rho_0}{\rho_2}\frac{1}{R^2h} \ll 1$. These assumptions respectively correspond to the vanishing crater deceleration and $\Delta\rho \ll \rho_2$ during the mixing stage. 
Using these assumptions, equations \ref{eq:mixing_conservation_adim} and \ref{eq:mixing_buoy_drag_adim} become
\begin{equation}
    \left\{
    \begin{array}{l}
        \dot{h}=-2\frac{\dot{R}}{R}h+u'\\
        \dot{u'}=-C\frac{u'^2}{h}
    \end{array}
    \right..
    \label{eq:h_equation_mixing}
\end{equation}
Assuming a $2/5$ power-law solution for $R$ (equation \ref{eq:energy_model_assum}), and using $h(1)=h_0$ and $\dot{h}(1)=\dot{h}_0$ as initial conditions, a solution to equation \ref{eq:h_equation_mixing} is
\begin{equation}
    h(t)=h_0\left[1+Q(t-1)\right]^{-4/5}\left\{1+A\left[\left[1+Q(t-1)\right]^{9/5}-1\right]\right\}^{\frac{1}{1+C}},
    \label{eq:h_power_mixing}
\end{equation}
where $A=\frac{1}{9}(C+1)(4+\frac{5}{Q}\frac{\dot{h}_0}{h_0})$ (figure \ref{fig:h_time_01}, dash-dotted lines).
The value $C=0.7$ required to fit experimental data is smaller than the value obtained by fitting experimental data with the full numerical model (equations \ref{eq:energy_model_assum}, \ref{eq:mixing_conservation_adim} and \ref{eq:mixing_buoy_drag_adim}). If the latter was used, the analytical solution would underestimate the layer thickness in the mixing stage since the value of $C$ depends strongly on the assumption made to obtain this solution, \textit{i.e.} approximated geometrical term, neglected buoyancy term, $\Delta\rho \ll \rho_2$, and $2/5$ power-law for $R$.

\subsubsection{Maximum mixing layer thickness}

\begin{figure}
    \centering
    \includegraphics[width=1\linewidth]{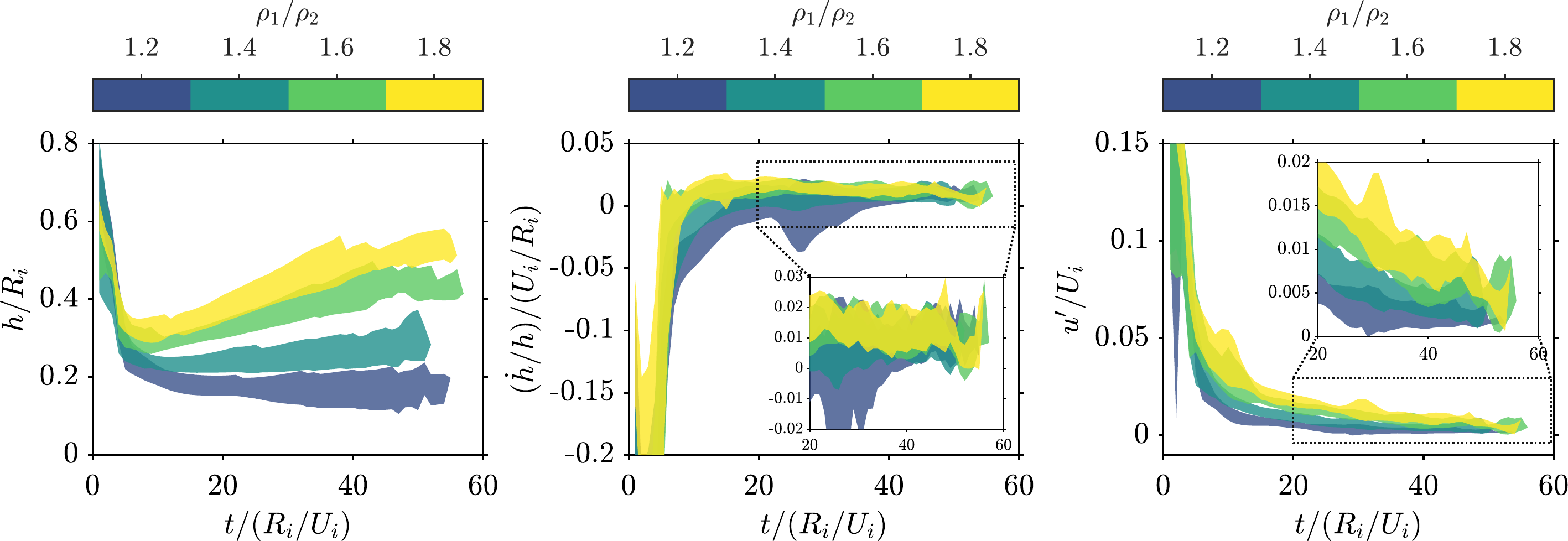}
    \caption{(a) Experimental mixing layer thickness $h$, normalised by the drop radius $R_i$, as a function of time, normalised by the drop free-fall time $R_i/U_i$.
    (b) Experimental mixing layer growth rate $\dot{h}/h$, normalised by the drop free-fall rate $U_i/R_i$, as a function of time, normalised by the drop free-fall time $R_i/U_i$.
    (c) Estimated inward flux density due to mixing $u'$ (from equation \ref{eq:mixing_conservation_adim}), normalised by the impact velocity $U_i$, as a function of time, normalised by the drop free-fall time $R_i/U_i$.
    Experiments are clustered by density ratio group, the extent of which is defined by the standard deviation of the experiments in that group. Colours scale as the density ratio $\rho_1/\rho_2$.}
    \label{fig:h_full_01}
\end{figure}

Figure \ref{fig:h_full_01} shows the mixing layer thickness $h$, growth rate $\dot{h}/h$, and the estimated mixing term $u'$ as a function of time, for all experiments, grouped by density ratio.
The maximum layer thickness increases significantly with the initial density ratio (figure \ref{fig:h_full_01}a).
In the mixing stage (typically $t/(R_i/U_i)>15$), an increased initial density difference promotes mixing by the RT instability, but leads only to a slight increase of the mixing layer growth rate (figure \ref{fig:h_full_01}b) and the estimated mixing term (figure \ref{fig:h_full_01}c).
The increase of the maximum layer thickness with the initial density ratio is in fact controlled by the time window available for the mixing layer to actually grow.
On one hand, the transition time $t_c$, \textit{i.e.} the time at which geometrical effects become comparable to RT mixing, corresponds to the lower limit of the available time window. Since $t_c$ decreases when the initial density ratio increases (equation \ref{eq:mixing_model_tc}), expanding in this way the time window, it is consistent with the increasing effect of the density ratio on the maximum layer thickness.
On the other hand, the maximum opening timescale $t_{max}$, \textit{i.e.} the time at which the crater radius reaches a maximum, corresponds to the upper limit of the available time window. $t_{max}$ being an increasing function of the density ratio (equation \ref{eq:energy_model_tmax}), it is also consistent with the increasing effect of the density ratio on the maximum thickness. Since $t_{max}$ also increases with the Froude number, the time window available for the mixing layer to develop, and consequently the maximum thickness, may also increase with the Froude number.

\begin{figure}
    \centering
    \includegraphics[width=1\linewidth]{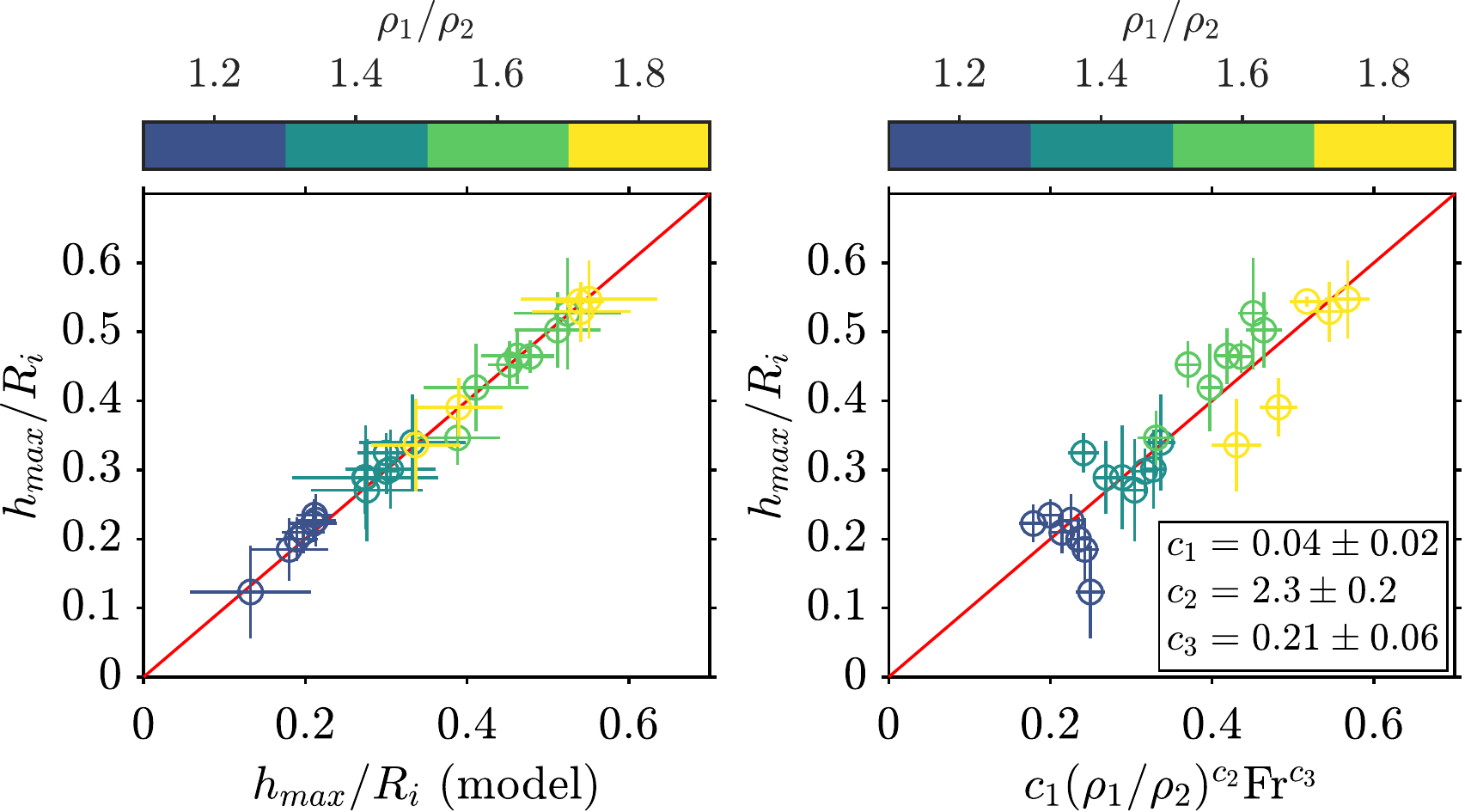}
    \caption{(a) Experimental maximum mixing layer thickness $h_{max}$, normalised by the drop radius $R_i$, as a function of the maximum thickness predicted by the mixing model (equations \ref{eq:energy_model_adim}, \ref{eq:mixing_conservation_adim} and \ref{eq:mixing_buoy_drag_adim}), normalised by the drop radius $R_i$.
    (b) Experimental maximum mixing layer thickness $h_{max}$, normalised by the drop radius $R_i$, as a function of the least-square best-fit power-law scaling $h_{max}^*$, using the density ratio $\rho_1/\rho_2$ and the Froude number $Fr$ (equation \ref{eq:mixing_model_hmax}). Colours scale as the density ratio $\rho_1/\rho_2$.}
    \label{fig:h_max_01}
\end{figure}

The experimental maximum mixing layer thickness is first compared with the maximum thickness obtained from the model (figure \ref{fig:h_max_01}a), leading to a good agreement. Since the maximum layer thickness is expected to depend on both the initial density ratio and the Froude number through power-laws, experimental data are then fitted using a power-law scaling $h_{max}^*$ (figure \ref{fig:h_max_01}b). A good agreement is obtained using
\begin{equation}
    h_{max}^*=c_1\left(\frac{\rho_1}{\rho_2}\right)^{c_2}Fr^{c_3},
    \label{eq:mixing_model_hmax}
\end{equation}
where $c_1=0.04\pm0.02$, $c_2=2.3\pm0.2$ and $c_3=0.21\pm0.06$.
This scaling is consistent with the qualitative observations of figure \ref{fig:RT_regime_01}, \textit{i.e.} a maximum mixing layer thickness increasing with the initial density ratio and the Froude number. Given the scalings for $t_c$ and $t_{max}$, it also agrees with an increased time window available for the mixing layer to develop.

\subsection{Early-time wavelength}

Experimental instability wavelengths at early times are converted into an equivalent spherical harmonic degree, and compared to a scaling law and to an approximate linear stability analysis model \citep{chandrasekhar_1955}.
From the number of plumes $n$, counted on the hemispherical section of the density interface at $t/(R_i/U_i)=10$, a typical instability wavelength $\lambda=\pi R/n$ is derived. The corresponding degree of maximum instability $l_{max}$ is then obtained using the Jeans relation \citep{jeans_1923}
\begin{equation}
    \sqrt{l_{max}(l_{max}+1)}=\frac{2 \pi R}{\lambda}.
    \label{eq:jeans_relation}
\end{equation}

\begin{figure}
    \centering
    \includegraphics[width=1\linewidth]{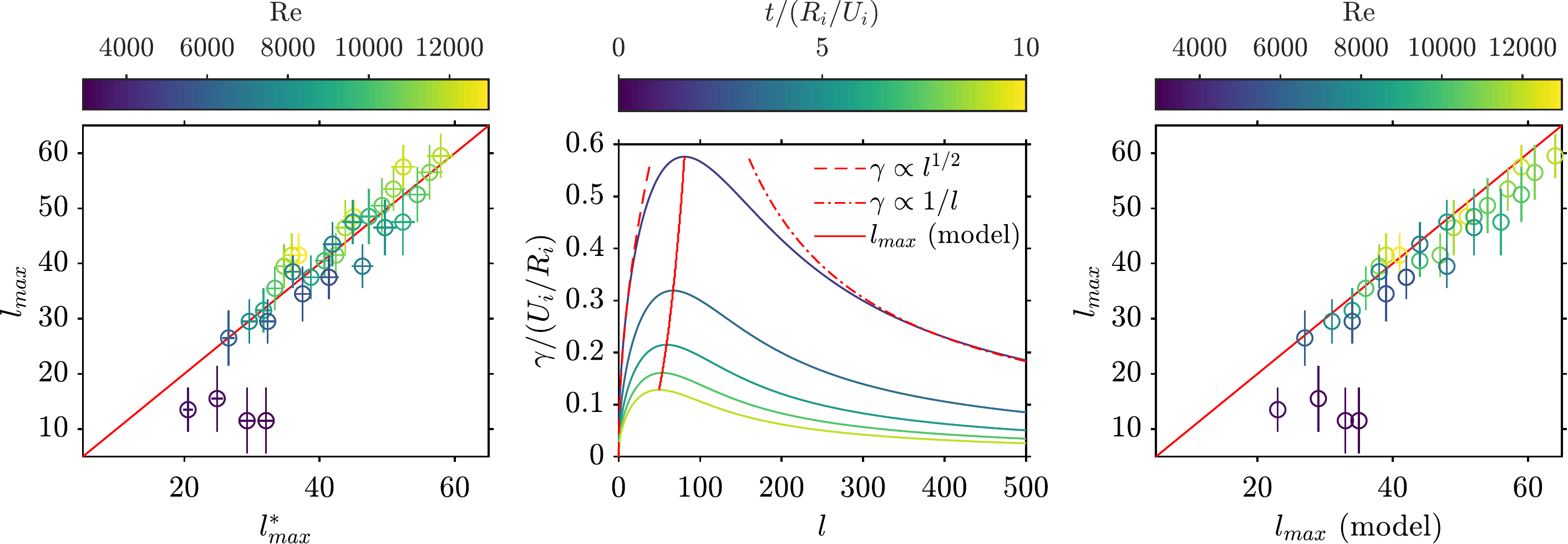}
    \caption{(a) Experimental harmonic degree of maximum instability $l_{max}$, as a function of the harmonic degree of maximum instability scaling $l_{max}^*$ (equation \ref{eq:wavelength_lmax}). Colours scale as the Reynolds number $Re$.
    (b) Theoretical growth rate of the instability (equation \ref{eq:wavelength_chandra_1}-\ref{eq:wavelength_chandra_2}), as a function of the harmonic degree $l$, at several times $t/(R_i/U_i)$. Dashed line and dash-dotted line correspond respectively to $l^{1/2}$ and $1/l$ scalings. The solid line corresponds to the theoretical degree of maximum instability.
    In this case, $\rho_1/\rho_2=1.8$, $\nu_1/\nu_2=1$, $Fr=10^3$, $Bo=0.75$, $Re=10^4$ $\phi=0.38$, $\xi=0.34$.
    (c) Experimental harmonic degree of maximum instability $l_{max}$, as a function of the theoretical degree of maximum instability (equations \ref{eq:wavelength_chandra_1}-\ref{eq:wavelength_chandra_2}). Colours scale as the Reynolds number $Re$.}
    \label{fig:wn_01}
\end{figure}

Concerning the scaling law, the instability wavelength is assumed to depend on the balance between an effective mixing layer acceleration $(\Delta\rho_0/\rho_2)\ddot{R}$ and the viscosity of the ambient fluid $\nu_2$, which respectively produce and damp the instability. The scaling for the wavelength is then $\lambda \sim \{\nu_2^2/[(\Delta\rho_0/\rho_2)\ddot{R}]\}^{1/3}$, which after nondimensionalization gives
\begin{equation}
    \lambda \sim \left(\frac{\Delta\rho_0}{\rho_2}\ddot{R}\right)^{-1/3} Re^{-2/3}.
    \label{eq:wavelength_lambda}
\end{equation}
Assuming that $\ddot{R}$ and $R$ respectively scale as $R_{max}/t_{max}^2$ and $R_{max}$ (see equations \ref{eq:energy_model_Rmax} and \ref{eq:energy_model_tmax}), and using the Jeans relation (equation \ref{eq:jeans_relation}), the scaling for the wavelength gives a scaling for the degree of maximum instability
\begin{equation}
    l_{max}^*=c~\phi^{1/4} \xi^{-1/3} Fr^{-1/12} \left(\frac{\rho_1}{\rho_2}\right)^{1/4} \left(\frac{\Delta\rho_0}{\rho_2}\right)^{1/3} Re^{2/3},
    \label{eq:wavelength_lmax}
\end{equation}
where $c=0.177\pm0.005$ is a least-squares best-fit prefactor obtained from experimental data (figure \ref{fig:wn_01}a).

Harmonic degrees of maximum instability are shown in figure \ref{fig:wn_01}a, for all experiments, as a function of their derived scaling $l_{max}^*$. Experimental data are mostly proportional to the scaling, except for Reynolds number smaller than 4000. In this case, the crater differs from the hemispherical shape and from the purely radial acceleration assumed in the scaling.

Experimental results are also compared to a linear stability analysis model \citep{chandrasekhar_1955}. An incompressible fluid sphere having a density $\rho_1$ and a viscosity $\mu_1$ is radially accelerated into an ambient fluid having a density $\rho_2$ and a viscosity $\mu_2$. The mean radius $R$ of the sphere is assumed to be constant with time.  Perturbations at the interface are expanded in spherical harmonics of degree $l$. Using a variational principle, an approximate dispersion relation for the growth rate $\gamma$ of the instability is obtained
\begin{equation}
    \gamma^2+A\left(l,\rho_1/\rho_2,\mu_1/\mu_2\right)\frac{\gamma}{R^2}\frac{1}{Re}-B\left(l,\rho_1/\rho_2\right)\frac{\ddot{R}}{R}=0,
    \label{eq:wavelength_chandra_1}
\end{equation}
with
\begin{equation}
    \left\{
    \begin{array}{l}
        A=\frac{2l(l+1)\left[l+(l+1)(\mu_1/\mu_2)\right]+(2l+1)^2(1-\mu_1/\mu_2)}{l+(l+1)(\rho_1/\rho_2)}\\
        B=\frac{l(l+1)(1-\rho_1/\rho_2)}{l+(l+1)(\rho_1/\rho_2)}
    \end{array}
    \right..
    \label{eq:wavelength_chandra_2}
\end{equation}
This dispersion relation, corresponding to a dimensionless version of equation 90 in \citet{chandrasekhar_1955}, is in agreement with exact results derived by \citet{chandrasekhar_1955}.

The growth rate of the instability is obtained by solving simultaneously the dispersion relation (equation \ref{eq:wavelength_chandra_1}) and the mean crater radius evolution (equation \ref{eq:energy_model_adim}). The time evolution of the theoretical growth rate as a function of the spherical harmonic degree is thus obtained for all experiments (\textit{e.g.} figure \ref{fig:wn_01}b).
The time evolution derives from the time dependent terms of equation \ref{eq:wavelength_chandra_1}, \textit{i.e.} $R(t)$ and $\ddot{R}(t)$. Since the model of \citet{chandrasekhar_1955} is a static model, \textit{i.e.} the average position of the density interface is constant with time, results should be considered carefully. The dynamic problem, \textit{i.e.} with a moving density interface, is much more challenging \citep[\textit{e.g.}][]{prosperetti_1977} and will not be addressed here.
At a given time, the instability growth rate reaches a maximum owing to viscosity effects, giving the theoretical degree of maximum instability $l_{max}$ (figure \ref{fig:wn_01}b). For small degrees, the instability growth rate scales as $l^{1/2}$, as expected from inviscid planar geometry cases \citep{rayleigh_1899,taylor_1950}. For large degrees, viscosity effects start to develop due to the larger velocity gradients involved. Viscosity dissipates short wavelength energy, leading to a $1/l$ decay of the instability growth rate, as in viscous planar geometry without surface tension \citep{chandrasekhar_1961}.

In order to compare the model with experimental data, the theoretical degree of maximum instability is calculated at $t/(R_i/U_i)=10$, \textit{i.e.} the time at which the experimental degree of maximum instability is measured. Figure \ref{fig:wn_01}c shows a close agreement between experimental data and the model, except for Reynolds number smaller than 4000, as previously explained.

\section{Geophysical implications}
\label{sec:geophysical}

After the impact, the metal core of the colliding body migrates toward the planetary core due to the density contrast with the surrounding silicates \citep{rubie_2015}. Part of the migration occurs in a fully molten magma ocean where the metal is expected to descend as a turbulent thermal and equilibrate with silicates \citep{deguen_2011,deguen_2014}. The metal phase then undergoes a vigorous stirring \citep{lherm_2018}, leading to its fragmentation \citep{landeau_2014,wacheul_2014,wacheul_2018} into centimetric drops \citep{stevenson_1990,karato_1997,rubie_2003,ichikawa_2010}.
However, these models assume that the metal cores are released as a compact volume in the magma ocean, which is not true after a planetary impact \citep{kendall_2016,landeau_2020}. Their initial conditions may be improved by considering the impact stage.

In order to estimate the mixing produced by the spherical RT instability during the opening stage of planetary impacts, a relevant quantity is the mass of ambient silicates that mixes with the impacting core during crater opening \citep{deguen_2014}. If an impactor with a radius $R_i$, a volume fraction of metal $f_m$, a metal core density $\rho_m$, and a silicate mantle density $\rho_s$, impacts a planetary target, the dimensionless mass of equilibrated silicates is $\Delta=M_s/M_m$, where $M_m=f_m\rho_m(4/3)\pi R_i^3$ is the mass of the metal core and $M_s=\rho_s[2\pi R_{max}^2h_{max}-(4/3)\pi R_i^3]$ is the mass of entrained silicates. After nondimensionalization, the mass of silicates mixed with metal gives
\begin{equation}
    \Delta=\frac{\rho_s}{\rho_m}\left(\frac{3}{2}\frac{1}{f_m}R_{max}^2h_{max}-1\right).
    \label{eq:delta}
\end{equation}

\begin{figure}
    \centering
    \includegraphics[width=\linewidth]{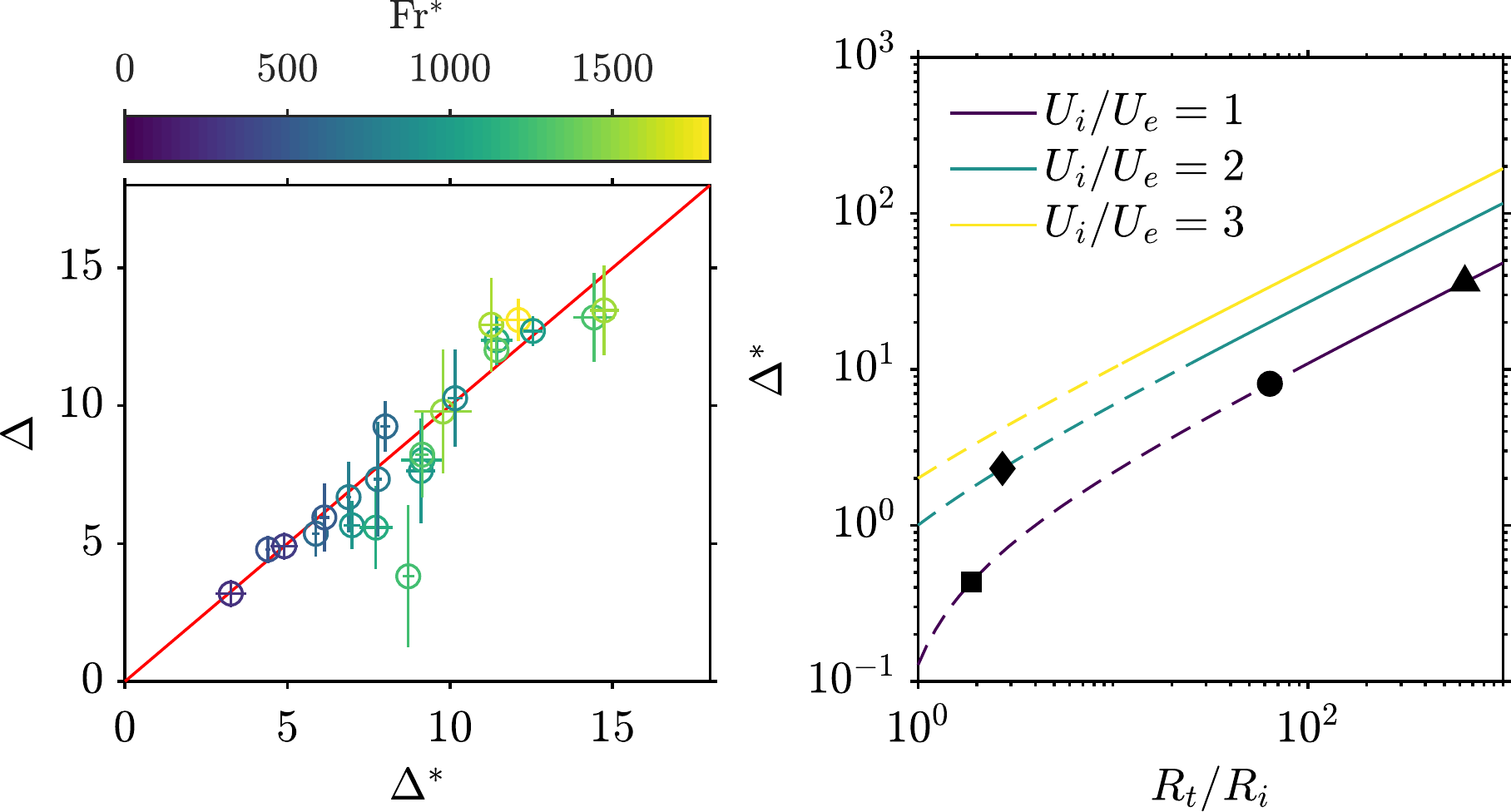}
    \caption{(a) Experimental dimensionless mixing mass $\Delta$, as a function of the dimensionless mixing mass scaling $\Delta^*$ (equation \ref{eq:delta_scaling}), using $f_m=1$. Colours scale as the modified Froude number $Fr^*$. (b) Dimensionless mixing mass scaling $\Delta^*$ as a function of the target to impactor radius $R_t/R_i$ (equation \ref{eq:Fr_planet}), for several impact velocities $U_i$, and using $f_m=0.16$ and $\rho_m/\rho_s=2$. Triangle: impactor of 10 km in radius onto a Earth-sized target. Circle: impactor of 100 km in radius onto a Earth-sized target. Square: canonical Moon-forming impact with a Mars-sized impactor \citep{canup_2004}. Diamond: fast-spinning Earth Moon-forming impact with a fast ($U_i=2U_e$) and small ($R_i/R_t=0.3$) impactor \citep{cuk_2012}. Dashed lines correspond to an extrapolated range of Froude number, \textit{i.e.} $Fr<200$, which is outside of the experimental Froude number range.}
    \label{fig:delta_01}
\end{figure}

Using scaling laws for $R_{max}$ (equation \ref{eq:energy_model_Rmax}), $h_{max}$ (equation \ref{eq:mixing_model_hmax}), and implicitly $\phi$ (equation \ref{eq:phi_scaling}), a scaling law for the mixing mass is obtained
\begin{equation}
    \Delta^*=\frac{\rho_s}{\rho_m}\left(\frac{1}{f_m}c_1\left(\frac{\bar{\rho}}{\rho_s}\right)^{c_2}Fr^{c_3}-1\right)
    \label{eq:delta_scaling}
\end{equation}
where $c_1=0.1\pm0.05$, $c_2=2.8\pm0.2$ and $c_3=0.63\pm0.06$. In this scaling, the density ratio is defined with $\bar{\rho}/\rho_s$, where $\bar{\rho}=\rho_mf_m+\rho_s(1-\rho_s)$ is the mean density ratio of the impactor, because it derives from the crater size and the maximum mixing layer thickness scalings (equations \ref{eq:energy_model_Rmax} and \ref{eq:mixing_model_hmax}, respectively), which indeed use the mean density of the impactor.
This scaling law is validated on experimental data in figure \ref{fig:delta_01}a, using $f_m=1$ since the drop is a one-phase fluid.

In the context of planetary impacts, the Froude number is given by
\begin{equation}
    Fr=2\frac{R_t}{R_i}\frac{U_i^2}{U_e^2}
    \label{eq:Fr_planet}
\end{equation}
where $U_e=\sqrt{2gR_t}$ is the escape velocity, and $R_t$ is the radius of the target planet.
The impact velocity of colliding bodies during accretion is typically one to three times the escape velocity \citep{agnor_1999,agnor_2004}, which means that the Froude number depends mainly on the target to impactor radius ratio. $Fr$ is about 1 for giant impacts, comparable in size to the target, but increases by several order of magnitude for small colliding bodies.

Using equations \ref{eq:delta_scaling} and \ref{eq:Fr_planet}, the estimated mass of silicates mixed with the impacting core during crater opening $\Delta^*$ is calculated as a function of the target to impactor radius ratio $R_t/R_i$ (figure \ref{fig:delta_01}b). We use $f_m=0.16$ and $\rho_m/\rho_s=2$ to match the internal structure of a realistic differentiated impactor \citep{canup_2004}.
Since the Froude number increases with the target to impactor radius ratio, it means that a smaller colliding body will produce more mixing, relative to their size, than giant impactors. For example, impactors with a 10 km and 100 km radius (figure \ref{fig:delta_01}b, triangle and circle, respectively) will then mix with 36.1 and 8.1 times its own mass, respectively.

Concerning the Moon-forming impact, the canonical impact scenario with a Mars-sized impactor \citep{canup_2004} is expected to mix with 0.4 times its own mass during this crater opening stage (figure \ref{fig:delta_01}b, square). However, the fast-spinning impact scenario associated with a faster ($U_i=2U_e$) and smaller ($R_i/R_t=0.05$) colliding body \citep{cuk_2012} is expected to mix with 2.3 times its own mass (figure \ref{fig:delta_01}b, diamond).
Recent experiments estimate the mass of equilibrated silicates during the impact, considering both the crater formation, its collapse into an upward jet, and the collapse of the jet \citep{landeau_2020}. In the same conditions, the 100 km radius impactor, the canonical Moon-forming impactor and the fast-spinning Earth impactor respectively mix with 168, 1.5 and 12 times the impactor mass. It corresponds to 4 to 20 times the mass mixed by the spherical RT instability, which is in agreement with an impact-induced mixing mostly dominated by the collapse of the jet \citep{landeau_2020}.
These giant impacts scenario involve small target to impactor radius, corresponding to an extrapolated range of Froude number, \textit{i.e.} $Fr<200$ (figure \ref{fig:delta_01}b, dashed lines), which is outside of the experimental Froude number range. The mass of equilibrated silicates extrapolated for large impactor thus has to be considered carefully. 

\section{Conclusion}
\label{sec:conclusion}

%CONCLUSION
A phenomenological description of impact cratering experiments has shown that crater deceleration after impact is responsible for a density-driven perturbation at the drop-pool interface, interpreted as a spherical Rayleigh-Taylor instability.
An energy conservation model for the crater radius evolution has been derived (equation \ref{eq:energy_model_adim}) and compared with backlight experiments, resulting in scaling laws for the maximum crater radius (equation \ref{eq:energy_model_Rmax}) and the opening timescale (equation \ref{eq:energy_model_tmax}).
In combination with this energy model, a mixing layer evolution model involving two stages has been derived (equations \ref{eq:mixing_conservation_adim} and \ref{eq:mixing_buoy_drag_adim}). The mixing layer dynamics is first controlled by the geometrical evolution of the crater, then by the balance between residual geometrical effects and mixing produced by the Rayleigh-Taylor instability. Scaling laws for the transition timescale between stages (equation \ref{eq:mixing_model_tc}) and the maximum mixing layer thickness (equation \ref{eq:mixing_model_hmax}) are obtained. The instability wavelength is also investigated using a an approximate linear stability analysis model (equations \ref{eq:wavelength_chandra_1}-\ref{eq:wavelength_chandra_2}), and a scaling law is obtained (equation \ref{eq:wavelength_lmax}).
These results have geophysical implications concerning the differentiation of terrestrial planets, in particular by estimating the mass of silicates that equilibrate with the core of the impactors during the impact of planetesimals on a magma ocean.

%PERSPECTIVES & LIMITATIONS
When experimental scaling laws are extrapolated to planetary conditions, we assume that physical processes observed in the experiments are self-similar at the planetary scale.
In order to validate these scaling laws, new experiments involving larger impactors are required to obtain Froude numbers relevant to giant impacts, and investigate the possible effect of the Reynolds number on the mixing layer.
Furthermore, several physical aspects neglected in our experiments have to be investigated experimentally or numerically, in order to examine their effect on cratering and mixing dynamics.
For giant impacts, geometrical effects related to the sphericity of both the impactor and the target, and to impact angle, may influence metal/silicate equilibration.
Immiscibility between metal and silicates may also change the mixing dynamics, in particular with the fragmentation of the metal phase. In this context, the viscosity and diffusivity contrasts may influence thermal and chemical transfers between phases.
Finally, since planetary impact velocities are larger than the sound velocity in silicates, the role of shock waves and compressibility effects, \textit{i.e.} the Mach number, on cratering and mixing scaling laws has to be investigated.\\

%ACKNOWLEDGEMENTS
This project has received funding from the European Research Council (ERC) under the European Unions Horizon 2020 research and innovation programme (grant agreement 716429).
We thank M. Moulin for the help with the design and construction of the experimental apparatus and J. Vatteville for the help with the imaging equipment. 

\bibliographystyle{jfm}
\bibliography{references}

\end{document}